\documentclass[acmtog]{acmart}

\usepackage{booktabs} 

\usepackage{listings}
\usepackage[export]{adjustbox}

\usepackage[ruled]{algorithm2e} 

\SetAlFnt{\small}
\SetAlCapFnt{\small}
\SetAlCapNameFnt{\small}
\SetAlCapHSkip{0pt}

\setcopyright{acmcopyright}

\usepackage{wrapfig}
\usepackage{soul}
\usepackage[normalem]{ulem}
\usepackage{amsmath}
\usepackage{tikz}
\usepackage{diagbox}
\usepackage{makecell}

\newcommand{\revision}[1]{\textcolor{black}{#1}}

\newcommand{\stkout}[1]{\revision{\ifmmode\text{\sout{\ensuremath{#1}}}\else\sout{#1}\fi}}

\hypersetup{draft}

\begin{document}
\title[]{Fast Tetrahedral Meshing in the Wild}

\author{Yixin Hu}
\affiliation{%
  \institution{New York University}
  \country{USA}
  }
\email{yixin.hu@nyu.edu}

\author{Teseo Schneider}
\affiliation{%
  \institution{New York University}
  \country{USA}
  }
\email{teseo.schneider@nyu.edu}

\author{Bolun Wang}
\affiliation{%
  \institution{Beihang University}
  \country{China}
  }
\email{wangbolun@buaa.edu.cn}

\author{Denis Zorin}
\affiliation{%
  \institution{New York University}
  \country{USA}
  }
\email{dzorin@cs.nyu.edu}

\author{Daniele Panozzo}
\affiliation{%
  \institution{New York University}
  \country{USA}
  }
\email{panozzo@nyu.edu}

\renewcommand\shortauthors{Hu, Y. et al}

\begin{abstract}
We propose a new tetrahedral meshing method, \textsc{fTetWild}, to convert triangle soups into high-quality tetrahedral meshes. 
Our method builds on the TetWild algorithm, 
replacing the rational triangle insertion with a new incremental approach to construct and optimize the output mesh, interleaving triangle insertion and mesh optimization. Our approach makes it possible to maintain a valid floating-point tetrahedral mesh at all algorithmic stages, eliminating the need for costly constructions with rational numbers used by TetWild, while maintaining full robustness and similar output quality. This allows us to improve on TetWild in two ways. First, our algorithm is significantly faster, with running time comparable to less robust Delaunay-based tetrahedralization algorithms. Second, our algorithm is guaranteed to produce a valid tetrahedral mesh with floating-point vertex coordinates, while TetWild produces a valid mesh with rational coordinates which is not guaranteed to be valid after floating-point conversion. As a trade-off, our algorithm no longer guarantees that all input triangles are present in the output mesh, but in practice, as confirmed by our tests on the Thingi10k dataset, the algorithm always succeeds in inserting all input triangles.   

\end{abstract}






\begin{teaserfigure}
	\centering
	\includegraphics[width=\linewidth]{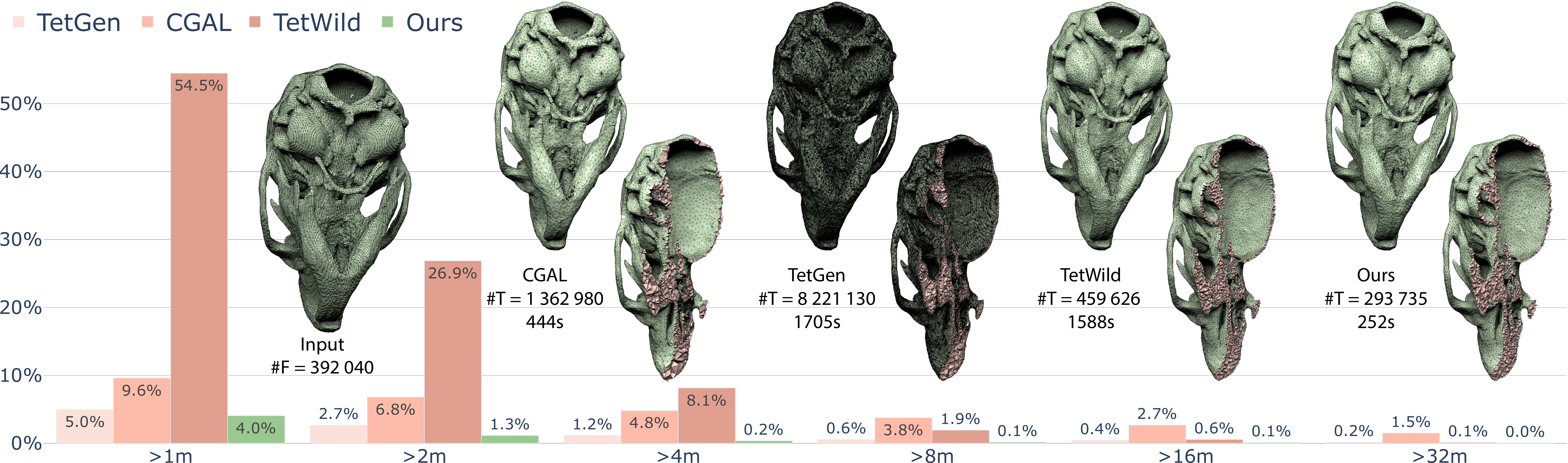}
	\caption{ The bar charts show the percentage of models requiring more than the indicated time for the different approaches over 4\,540 inputs (the subset of Thingi10k where all 4 compared algorithms succeed). Our algorithm successfully meshes \revision{98.7\%} of the input models in less than 2 minutes, and processes all models within 32 minutes. The comparison has been done using the experimental setup of TetWild~\cite{Hu:2018} and selecting a similar target resolution for all methods. The CGAL surface approximation parameter has been selected to be comparable to the envelope size used for TetWild and for our method.  The images above the plot show a mouse skull model (from micro-CT) tetrahedralized with \textsc{fTeWild} (right) compared with other popular tetrahedral meshing algorithms.
	}
	\label{fig:teaser}
\end{teaserfigure}

\maketitle

\section{Introduction}
Tetrahedral meshes are commonly used in graphics and engineering applications. \revision{Tetrahedral meshing algorithms usually take a 3D surface triangle mesh as input and output a volumetric tetrahedral mesh filling the volume bounded by the input mesh. Traditional tetrahedral meshing algorithms have strong assumptions on the input, requiring it to be a closed manifold, free of self-intersections and numerical unstably close elements, and so on. However, those assumptions often do not hold on imperfect 3D geometric data in the wild.}

The recently proposed Tetrahedral Meshing in the Wild (TetWild) \cite{Hu:2018} algorithm makes it possible to reliably tetrahedralize triangle soups by combining exact rational computations with a geometric tolerance to automatically address self-intersections, gaps and other imperfections in the input. The algorithm \revision{imposes no formal assumptions on the input mesh} and is extremely robust, opening the door to automatic processing and repair of large collections of 3D models.

However, {TetWild} has two downsides, one theoretical and one practical. The theoretical downside is that it does not guarantee the generation of a floating point tetrahedral mesh: the algorithm internally uses rational numbers, which are then converted to floating point in the process of mesh optimization.  While quite unlikely, it is possible that the mesh optimization stage will be unable to round all coordinates of the output mesh to floating point. The practical downside is the long running time compared with Delaunay-based tetrahedralization algorithms.

We introduce \textsc{fTetWild}, a variant of the TetWild algorithm addressing both these limitations \revision{while keeping the important properties of TetWild: robustness to imperfect input and ability to batch process large collections of models without parameter tuning, while producing high-quality tetrahedral meshes}. 
\revision{Differently from TetWild, which generates a polyhedral rational mesh inserting all triangles at once, we start from a floating point tetrahedral mesh, insert one input triangle at a time and re-tetrahedralize locally, rejecting the operations producing inverted or degenerate elements.} 
We then improve the quality of the mesh iteratively, and attempt to insert the rejected triangles into a higher quality mesh, 
\revision{which is less likely to fail.} 

\revision{Our algorithm always guarantees to generate a valid tetrahedral mesh with floating point vertex positions, independently from the stopping criteria or quality of the mesh. It might fail to insert few input triangles leading to a ``less accurate'' boundary preservation, however we never observe this behavior in our experiments.} The new algorithm can be implemented using floating point constructions, avoiding the 
\revision{overhead} 
associated with rational numbers. The use of floating point numbers also simplifies parallelization, which we use during mesh optimization to further improve the running time on large models.
\revision{Consequently, our new algorithm is significantly faster than TetWild, with running times comparable to Delaunay-based algorithms (Figure \ref{fig:teaser}), while providing the stronger guarantee of always producing a valid floating point output} at the same time. 

 These improvements make \textsc{fTetWild} more practical than TetWild not only for volumetric meshing problems, but also for mesh repair and approximate mesh arrangements. By combining \textsc{fTetWild} and some elements of \cite{Zhou:2016}, we obtain an approximate mesh arrangement algorithm for input triangle soups guaranteed to produce a valid floating point output. In comparison, the original algorithm presented in \cite{Zhou:2016} may fail to produce a floating-point output due to impossibility of rounding after the rational-arithmetic arrangement computation. 
 
 We demonstrate the robustness and practical utility of our algorithm by computing tetrahedral meshes on the Thingi10k dataset (10\,000 models) and computing approximate Booleans.
 We use the generated tetrahedral meshes to solve elasticity, fluid flow, and heat diffusion equations on complex geometric domains. The complete implementation of \textsc{fTetWild} is provided in the additional material, together with scripts to reproduce all results in the paper. 
\section{Related Work}
\label{sec:related}

We briefly review the literature on tetrahedral meshing (Section~\ref{sec:rel:meshing}), with an emphasis on envelope-based techniques, and we refer to \cite{Cheng:2012:DMG,Shewchuk2012} for a more detailed overview of the topic. \revision{We also review mesh repair and mesh arrangement algorithms} (Section~\ref{sec:rel:application}), since our technique can be also used in these settings to enable processing of imperfect geometry. 

\subsection{Tetrahedral Meshing}
\label{sec:rel:meshing}

\paragraph{Delaunay Meshing.}

The most studied and most widely used algorithms to generate tetrahedral meshes are based on the Delaunay condition~\cite{Chew:1993:GQM,Shewchuk:1998:TMG,Ruppert:1995:ADR,Sheehy:2012:NBO,Remacle:2017:ATL,Du:2003:TMG,Alliez:2005:VTM,Tournois:2009:IDR,MURPHY:2001:APP,CohenSteiner:2002:CDT,Chew:1987:CDT,Si:2005:MPL,Shewchuk:2002:CDT,Si:2014:ICA,Cheng:2008:APD,Boissonnat:2005:PGS,Jamin:2015:CAG,Dey:2008:DAD,Chen:2004:ODT,SHEWCHUK-triangle1,Aurenhammer1991,Shewchuk99lecturenotes,Cheng:2012:DMG,Aurenhammer2013,Bishop2016,Busaryev:RMI:2009,triangulation_in_cgal,Si:2015,George:2003,Weatherill:1994}. These methods are efficient and are widely used in commercial software. They can be applied to either point clouds inputs, or to tessellate the interior of manifold non self-intersecting meshes with no degenerate faces, but are not designed to deal with imperfect input, and, as a consequence, these techniques fail on  a significant fraction of data \emph{in the wild}~\cite{Hu:2018}. \revision{We provide a direct comparison with TetGen \cite{Si:2015}, the most commonly used code in this category, which builds on the techniques developed in \cite{George:2003,Weatherill:1994}, and CGAL \cite{cgal:wfzh-a2-18b} in Section \ref{sec:results}.}

\paragraph{Grid Methods.}
An alternative approach is the use of a background grid as a starting point~\cite{Yerry1983,Baker1988,BERN1994,Molino:2003:TMG,Bronson:2013:LCC,Labelle:2007:ISF,Doran:2013:ISI,code:quartet}. These algorithms fill the entire bounding box of the input with a regular lattice or with a hierarchical space partitioning, optionally intersect the background mesh with the input surface, and then discard the elements outside of the input. These methods are simpler and more robust than Delaunay methods, but still struggle with imperfect input geometry, and create high-quality elements only in the interior of the mesh, where the background mesh is preserved exactly. However, placing badly shaped triangles on the boundary is problematic for many applications. Our algorithm borrows the idea of a background mesh from these methods, but inserts the elements incrementally, interleaving mesh optimization stages to ensure that the final quality of the mesh is uniformly high. 

\paragraph{Front-Advancing Methods.} 
Another family of methods starts from the boundary, and inserts one element at a time, growing the volumetric mesh (i.e. marching in space), until the entire volume is filled \cite{George1971,Sadek1980,Peraire1987,Cuilliere:2013:ADM,Alauzet:2014:ACA,Haimes:2014:MMO}. These methods create high quality elements close to the boundary, but introduce many corner cases in the interior regions where the fronts meet, lowering the quality of the elements and making a robust implementation challenging.

\paragraph{Envelope Meshing.}

All methods discussed above assume a valid, manifold, non self-intersecting boundary input mesh, and are not designed to handle the imperfections which are common in real-world CAD and scanned data. This issue has been tackled for surface meshes in~\citet{Mandad:2015} by creating a surface approximation within a tolerance volume using a modified Delaunay refinement process. A similar idea has been exploited for volumetric meshing in TetWild~\cite{Hu:2018}, and its 2D counterpart TriWild \cite{Hu:2019}. The main idea of this work is to combine exact computation, using a hybrid kernel similar to \cite{Attene:2017}, and a surface envelope \cite{Hu:2017:EB}, which allows the resulting mesh to approximate the input instead of reproducing it exactly. Our method closely follows \cite{Hu:2018}, but we design our algorithm to avoid the use of exact computation. We compare the two techniques in Section \ref{sec:results}.

\paragraph{Mesh Improvement.} 
Many algorithms have been proposed to improve the quality of an existing tetrahedral mesh by displacing vertices or changing the local connectivity \cite{Canann1996,CANANN1993185,Lori1998,Lipman:2012:BDM,Chen:2004:ODT,Alliez:2005,FTB:2016:MVR,Feng:2018:COD,Hu:2018,Alexa:2019}.
Our method relies on the algorithm proposed in \citet{Hu:2018}, which uses a set of local operations to optimize the conformal AMIPS energy  \cite{Fu:2015,Rabinovich:2017}. We parallelized some of the steps of that algorithm (Section \ref{sec:method}), which is easier in our case since we only have floating point coordinates. 

\subsection{Applications}
\label{sec:rel:application}

\paragraph{Mesh Repair.}

Since our algorithm can be used for mesh repair, we review the most recent works on this topic, and we refer to \cite{Attene:2013} for a complete overview.

MeshFix \cite{Attene:meshfix:2010,Attene:2014:DRS} detects problematic regions in triangle meshes, and uses a set of local operations to heal them. The tool is very effective, but due to its use of a greedy algorithm it might delete large parts on a mesh.
The most recent mesh repair technique has been introduced in \cite{Hu:2018}: the algorithm generates a tetrahedral mesh and discards the generated tetrahedra, only keeping the boundary surface. While simple and effective, this techniques is computationally expensive, and thus only usable in batch processing applications. Our algorithm can be used in the same way, but its higher efficiency makes it more practical. \revision{We also propose a simple modification to the surface mesh extraction procedure to guarantee a manifold output.}

\paragraph{Booleans and Mesh Arrangements.}
Many approaches to performing Boolean operations on meshes were proposed, with some methods emphasizing robustness, other methods aiming to produce exact results, and another set prioritizing performance. In most cases, non-trivial assumptions are made on the input meshes: most commonly, these are required to be closed; in other cases, no self-intersections are allowed, or most restrictively vertices may be assumed in general position. 

CGAL, one of the most robust implementations of Boolean operations available \cite{Granados03booleanoperations}, relies on exact arithmetic, and uses a very general structure of Nef polyhedra \cite{Bieri:1988:ESO} to represent shapes. This allows one to obtain exact Boolean results in degenerate cases (e.g., when the result is a line or a point).  At the same time, the assumptions on the input are quite restrictive: the surfaces need to be closed and manifold (although the latter constraint could be eliminated). 

Another approach to achieve robustness at the expense of accuracy, is to convert input meshes to level sets  e.g. by sampling a signed distance function for each object \cite{museth2002level} and perform all operations on the level set functions. The obvious disadvantage of these methods is that their accuracy is limited by the resolution of the grid; the original mesh geometry is lost, and it is non-trivial to maintain even explicitly tagged features. These downsides are partially addressed by adaptive \cite{varadhan2004topology} and hybrid 
\cite{Pavic:2010,Wang11,zhao2011parallel}, the latter preserving 
mesh geometry away from intersections. All these methods rely on well-defined signed distance function, i.e., assume that input meshes are closed, and may still significantly alter the input geometry near intersections. \cite{schmidt2010meshmixer} does not use a signed distance function, but resembles these methods, in that it removes existing geometry near intersections and replaces it by new mesh connecting the two objects and approximating the result of the Boolean.  Binary Space Partitioning (BSP) based methods, starting from \cite{thibault1987set,naylor1990merging} are closest in their approach to ours. Using BSP trees preserves the input more accurately, and, along the way, creates a volume partition. However, it is prone to errors due to numerical instability of intersection calculations, and, due to global intersections of triangle planes, performs excessive refinement. 
\cite{Bernstein:2009:FEL} addresses the issue of non-robustness by using exact predicates, and  \cite{Campen:2010fr} reduces refinement by creating localized BSP trees in an octree.
Examples of highly efficient but non-robust software for computing Booleans are
\cite{Douze:2015}, \cite{Barki2015}, and \cite{Bernstein:Cork:2013}. A general position assumption is often required explicitly or implicitly.  In \cite{Zhou:2016} a robust way to compute \emph{mesh arrangements} is introduced, \revision{with Boolean operations as an application}. Robustness is achieved by using rational numbers for critical computations. To perform Booleans the mesh is required to be \revision{Positive Winding Number (PWN)}, which does not always hold in meshes in the wild \cite{Zhou:2016:TKA}. 

\citet{sheng2018efficient,sheng2018accelerated} use a combination of plane-based and vertex-based representations of mesh faces to improve robustness of basic operations needed for Boolean operations performed in floats. Their method achieves very high efficiency, at the expense of somewhat lower robustness compared to the state of the art \cite{Zhou:2016,Granados03booleanoperations}. Their method assumes that the input meshes enclose solids and are free of self-intersections.   \cite{magalhaes2017fast}  is an efficient technique using simulation-of-simplicity techniques to handle general intersections between objects, self-intersections or holes are not handled.  \cite{paoluzzi2017regularized} considers a general problem of arrangements of complexes in 2D and 3D, presenting a theoretical general merge algorithm, but do not consider the questions of robustness and handling imperfect inputs. 

Compared to existing methods, the application of \textsc{fTetWild} to Boolean operations is more conservative, in terms of mesh geometry changes and refinement, compared to level set and BSP-based methods, while maintaining their level of robustness.  At the same time, thanks to the geometric tolerance, \textsc{fTetWild} is capable of eliminating near-degenerate or overly refined triangles in the input model, which \cite{Zhou:2016} cannot do. We also make fewer assumptions on the inputs, allowing gaps, self-intersections, and degeneracies.

\begin{figure}
    \centering\footnotesize
    \includegraphics[width=\linewidth]{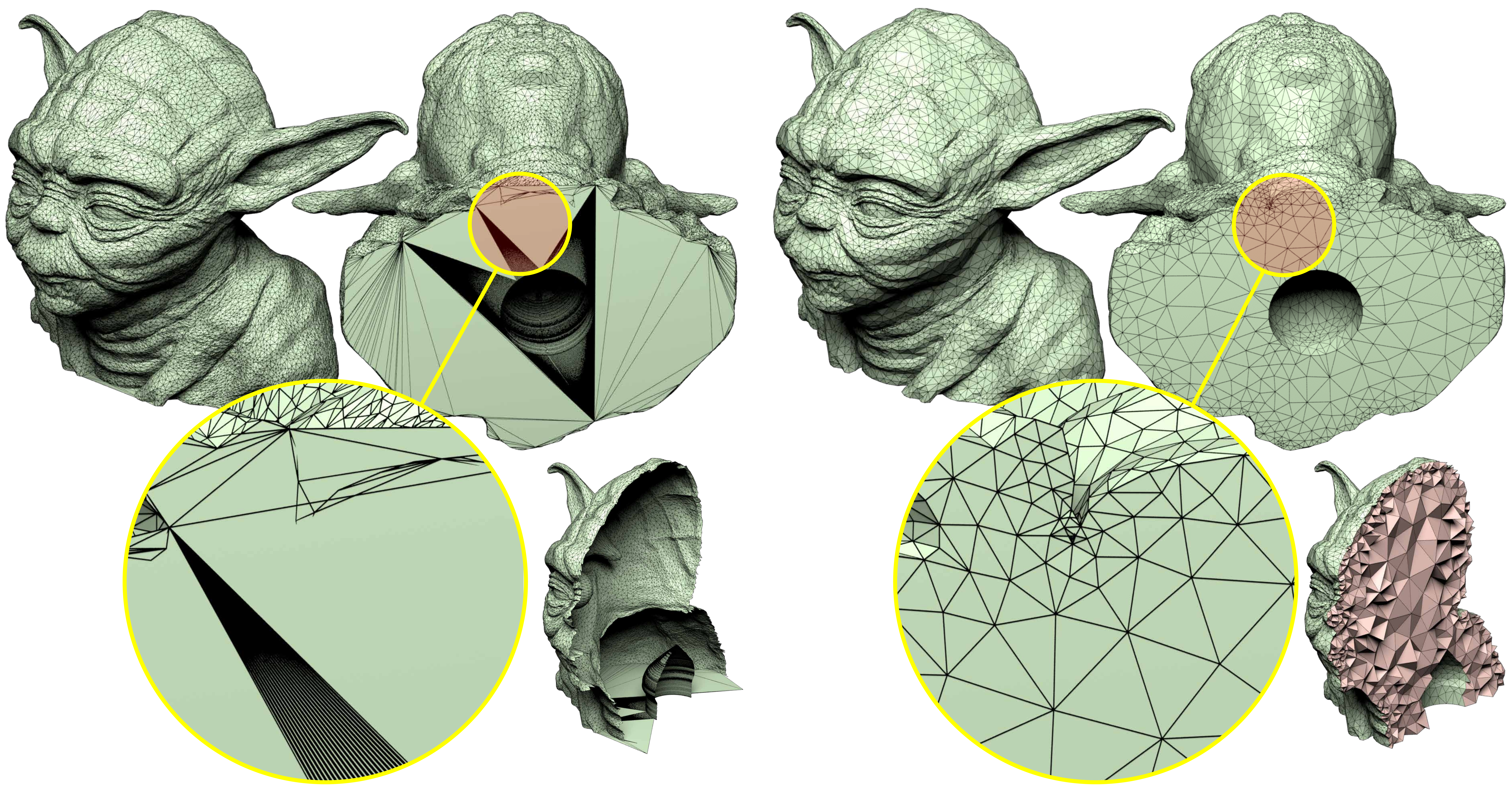}\\[0.4em]
    \parbox{.49\linewidth}{\centering Input\\\#F = 89\,466}\hfill
    \parbox{.49\linewidth}{\centering Our output 63s\\\#T = 92\,808\\Max energy = 8.0}\par
    \caption{\revision{Example of an input surface mesh with self-intersections and a bad triangulation on the base. \textsc{fTetWild} converts this model into a high-quality tetrahedral mesh.}}
    \label{fig:artifacts}
\end{figure}

\begin{figure*}
    \centering\footnotesize
    \includegraphics[width=\linewidth]{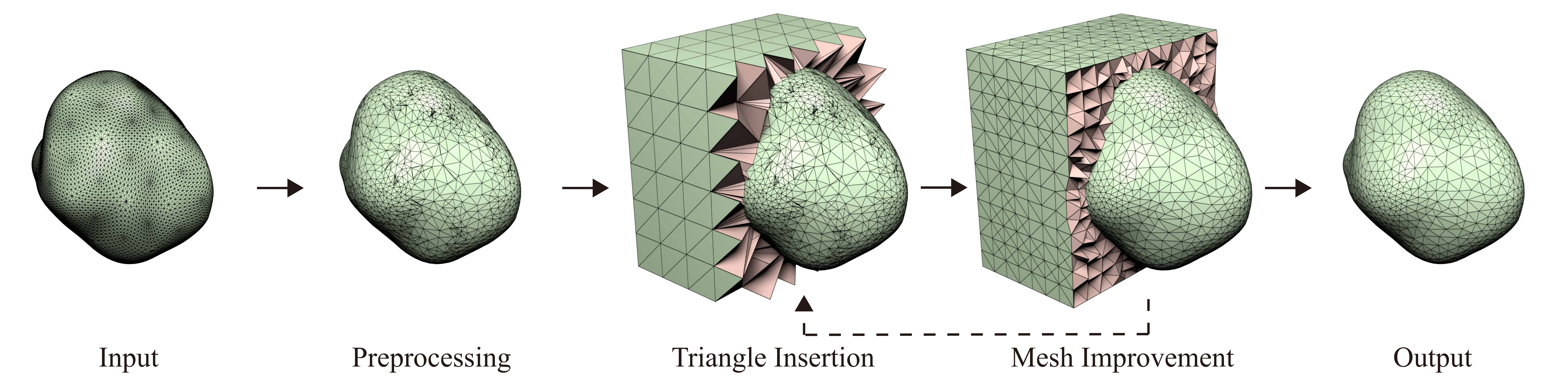}
    \caption{Overview of our algorithm. From left to right, the input mesh is simplified, a background mesh is created and the input faces are inserted, the mesh quality is optimized, and the final result is obtained by filtering the elements lying outside the input surface.}
    \label{fig:overview}
\end{figure*}

\section{Method}
\label{sec:method}
\revision{\textsc{fTetWild} takes as input a 3D triangle soup (i.e., a set of arbitrarily connected, potentially intersecting triangles with vertices potentially duplicated) whose vertices are represented in floating-point coordinates, representing the surface of an object. \revision{The algorithm has two user-defined parameters: target edge length $\ell$, and envelope size $\epsilon$. The $\epsilon$-envelope represents the maximal deviation from the input surface the user is willing to accept. For instance, in additive manufacturing applications $\epsilon$ can be the machining precision.}
It outputs a volumetric tetrahedral mesh, with floating-point vertex coordinates, whose elements are (1) non-inverted (i.e., positive volume) and (2) with boundary faces conforming to the input soup within a user-defined $\epsilon$-envelope. 
\textsc{fTetWild} makes \emph{no assumptions} on the input triangle soup and it is robust when handling imperfect input with self-intersections or small gaps. This robustness is achieved by allowing the faces of the tetrahedral mesh corresponding to the input surface to move inside an $\epsilon$-envelope (up to $\epsilon$ far from the input): self-intersections, degenerate and near-degenerate faces and gaps contained in the envelope are automatically removed when combined with proper mesh improvement operations (Figure~\ref{fig:artifacts}).}

\revision{
\paragraph{Similarities and Differences to Existing Face Insertion Algorithms.} The main challenge tackled in many existing tetrahedral meshing algorithm is the preservation of the input faces, which can be exact or approximate.
One of the best known algorithms exactly preserving the input faces is \cite{George:2003}, which proposed to subdivide a background mesh by intersecting it with input faces. This procedure can, however, introduce inverted elements due to floating-point rounding, which then need to be untangled, a difficult task for which no robust algorithm currently exists. A robust solution is proposed in TetWild~\cite{Hu:2018}, that initially inserts the faces exactly using rational numbers to avoid numerical problems, but is then forced to allow them to move to round the rational coordinates back to floating point. Although robust and conservative, this solution relies on expensive rational constructions, and it is not guaranteed to succeed in the rounding phase.}

\revision{Our method follows an approach similar to TetWild (see Appendix~\ref{app:tetwild} for a brief description of the algorithm), enabling small and controlled deviations from the input surface, but sidesteps the need for constructing a rational mesh, always using floating-point coordinates, while inheriting the robustness of TetWild. Algorithmically, there are three major differences: 
\begin{enumerate}
    \item \textsc{fTetWild} preserves the input faces by inserting one input triangle at a time into an existing background tetrahedral mesh. To facilitate the insertion it relaxes the insertion with a snapping tolerance (relatively larger than floating point machine precision) which is only possible thanks to the $\epsilon$-envelope.
    \item \textsc{fTetWild} always tetrahedralizes the region affected by the newly inserted face by looking up a pre-computed table and always maintains a valid inversion-free tetrahedral mesh (using exact predicates).
    \item \textsc{fTetWild} represents the vertices using only floating point coordinates, reducing the running time and memory consumption.
\end{enumerate}
}

\revision{We note that inserting a triangle might fail due to limitations of the floating-point representation. For instance, the inserted face can be arbitrarily close to one of the existing vertices and the insertion will introduce a tetrahedron with a volume numerically equal to zero. In this scenario, we rollback the problematic operation, mark the problematic face as un-inserted, iteratively perform mesh improvement operations on the whole mesh, and try to insert the face again when the mesh quality has increased. This procedure shows the only possible failure of \textsc{fTetWild}: the impossibility of adding some input faces. While this is indeed possible, it never manifested in our experiments. Note that even if some faces could not be inserted, \textsc{fTetWild} still outputs a valid mesh conforming to all other faces.}

\subsection{Algorithm Overview}

Our algorithm consists of four phases (Figure \ref{fig:overview}): (1) the input triangle soup is simplified while ensuring it stays in the $\epsilon$-envelope of the input (Section \ref{sec:met:preprocessing}), (2) a background mesh is generated and the triangles are iteratively inserted into it, if \revision{the insertion does not introduce inverted elements} (Section \ref{sec:met:insertion}), (3) the mesh is improved using local operations (Section \ref{sec:met:improvement}) and at the end of every three improvement iteration we re-attempt the insertion of input triangles that could not be inserted at phase 2, (4) the mesh elements are optionally filtered to remove the elements outside the surface or to perform Boolean operations (Section \ref{sec:met:filtering}).

During the whole procedure we ensure that the tetrahedral mesh remains \emph{valid}, that is, we ensure that (1) each element has positive volume (checked using exact predicates \cite{shewchuk97a,geogram}) and (2) all successfully inserted triangles, from now on called the \emph{tracked surface}, stay inside the $\epsilon$-envelope of the input.

Throughout the algorithm, we consider a distance between two points zero if it is below a numerical tolerance $\epsilon_\mathrm{zero}$. Similarly, we use  $\epsilon_\mathrm{zero}^2$, $\epsilon_\mathrm{zero}^3$ for areas and volumes respectively. We found that the performance of the algorithm are not heavily affected by this tolerance, as long as $\epsilon_\mathrm{zero} > 10^{-20}$: in our experiments we used $\epsilon_\mathrm{zero}=10^{-8}$.

\subsection{Envelope}
\revision{We use the envelope definition and the algorithms introduced in \cite{Hu:2018} to build the envelope and check if a triangle is contained in it. In particular, testing if a triangle is contained within the envelope is done by sampling the input triangle and checking if the samples are all within a slightly smaller envelope with the sampling error conservatively compensated \cite{Hu:2018}.}

\subsection{Preprocessing}
\label{sec:met:preprocessing}

We use the same preprocessing procedure proposed in \cite{Hu:2018} for simplifying the input: we merge vertices closer than $\epsilon_\mathrm{zero}$ and collapse an edge if: (1) \revision{it is a manifold edge (has no more than two incident triangles) and  vertex-adjacent edges are also manifold}, and (2) the collapsing operation does not move triangles outside \revision{a \emph{smaller} envelope of size $\epsilon_{\mathrm{prep}}<{\epsilon}$.
At this stage, we use $\epsilon_{\mathrm{prep}}=0.8{\epsilon}$ since this value gives space for snapping in triangle insertion (Section~\ref{sec:met:insertion}), and prevents vertices to be too close to the boundary of the envelope, thus leaving more freedom for surface vertices to move in the mesh improvement stage (Section~\ref{sec:met:improvement})}. \revision{On our dataset, we observed that changing this parameter has a minor impact on the running time and negligible effect on the output when in the range 0.7 to 0.999. Note that it cannot be set to 1 because it will prevent snapping (Section \ref{sec:met:insertion}). We use the value 0.8 since it is far from the bounds of this range.}


\begin{figure}
    \centering\footnotesize
    \includegraphics[width=.24\linewidth]{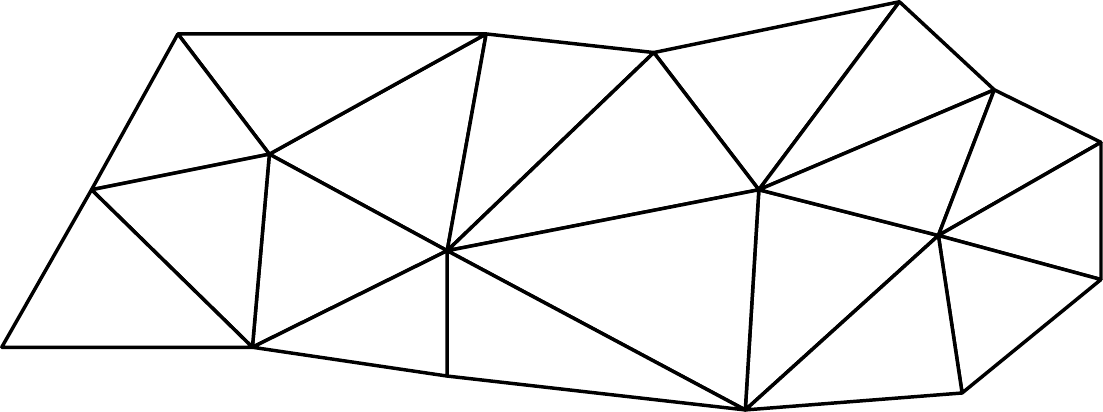}\hfill
    \includegraphics[width=.24\linewidth]{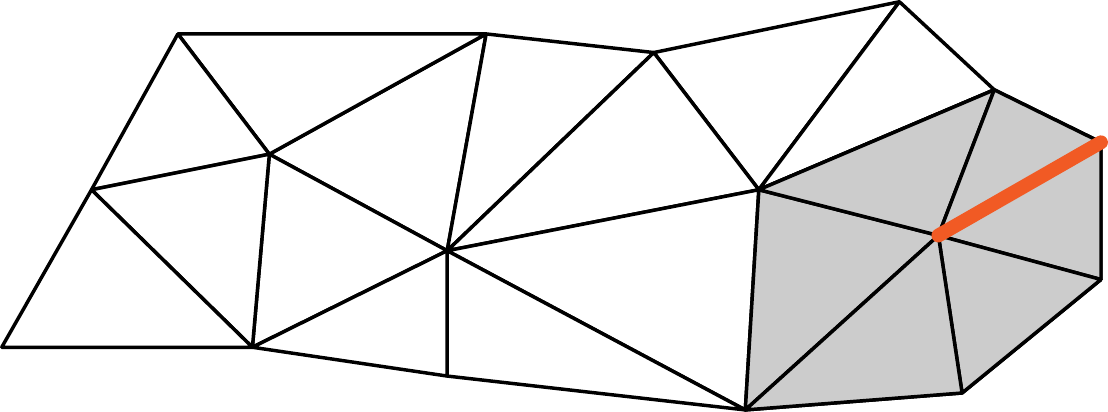}\hfill
    \includegraphics[width=.24\linewidth]{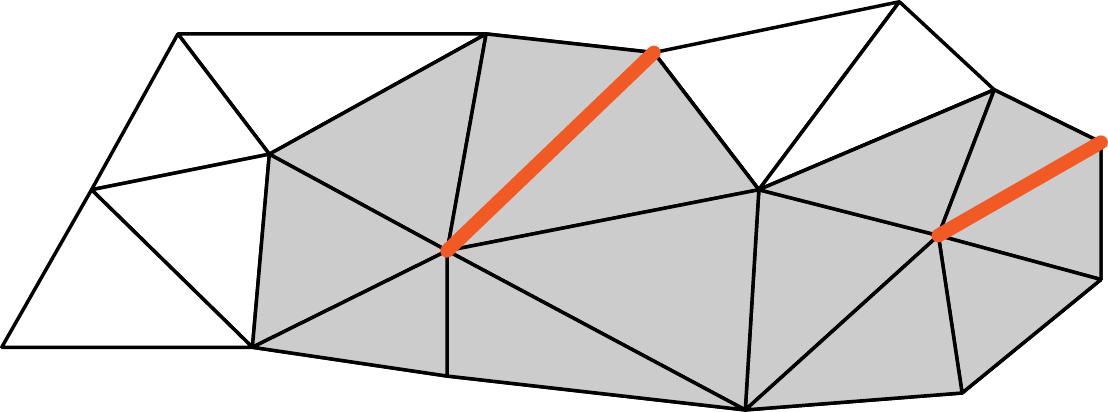}\hfill
    \includegraphics[width=.24\linewidth]{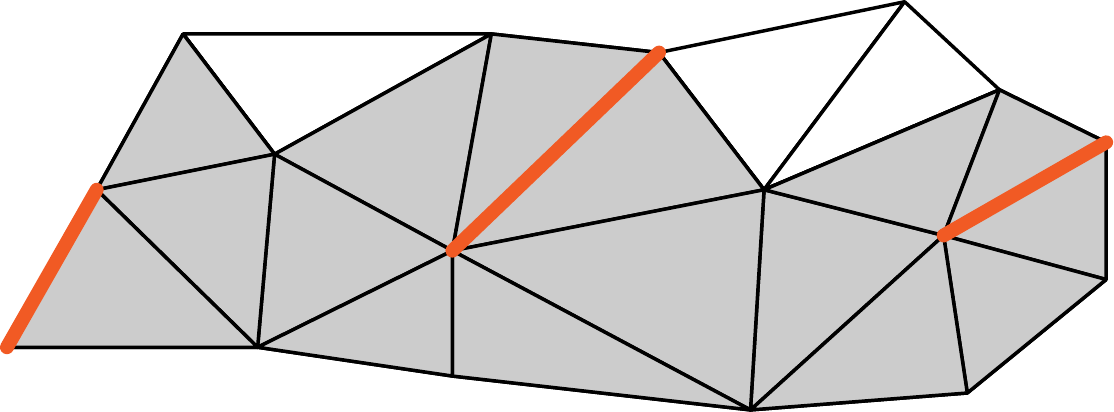}\hfill
    \caption{\revision{A two-dimensional example of edge coloring. From left to right: one parallel-independent edge is selected (in red), its vertex-adjacent triangles are colored in black. The algorithm proceeds until no edges can be marked (last image). In the end, all red edges can be safely collapsed in parallel without affecting other red edges.}}
    \label{fig:coloring}
\end{figure}

\revision{
Since the preprocessing step is computationally expensive, due to the envelope containment checks, we propose a basic parallelization strategy which leads, on average, to a 4x speedup of the preprocessing step when using 8 cores. Our parallel edge collapsing procedure uses a serial 2-coloring pass over all input faces. We mark all input triangles white in the initial stage. Then, iteratively, we mark all edges as \emph{parallel-independent} if all its vertex-adjacent triangles are white, and then  mark these triangles black (Figure~\ref{fig:coloring}). At this point, we can safely collapse all marked parallel-independent edges in parallel. We iterate this procedure until we are unable to remove more than $0.01\%$ of the original input vertices.}

\subsection{Incremental Triangle Insertion}
\label{sec:met:insertion}

\subsubsection{Background Mesh Generation}
\label{sec:met:bg-mesh}
\revision{
The triangle insertion stage requires a background mesh (which does not necessarily conform to the input triangles) which we create using Delaunay tetrahedralization~\cite{geogram} on the vertices from the preprocessing stage.  
Since we allow the surface to move within an $\epsilon$-envelope, we generate the background mesh for a bounding box $2\epsilon$ larger than the bounding box of the input vertices. Similarly to TetWild, additional points are added uniformly inside the box and at least $\epsilon$ away from the input faces before Delaunay tetrahedralization to obtain more uniformly-shaped initial elements. More precisely, the additional points are added in a regular grid with spacing of $d/20$ (where $d$ is the diagonal of bounding box of the input mesh), skipping inserting the additional points with distance to the input faces smaller than $\epsilon$. }

\subsubsection{Single Triangle Insertion}
\begin{figure*}
    \centering\footnotesize
    \includegraphics[width=\linewidth]{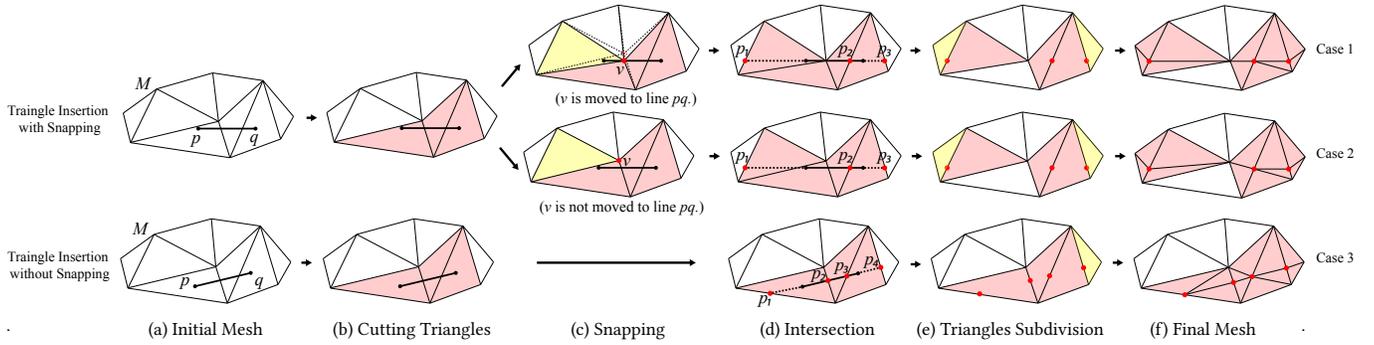}\par
    \parbox{.07\linewidth}{.}\hfill
    \parbox{.15\linewidth}{\centering (a) Initial Mesh}\hfill
    \parbox{.15\linewidth}{\centering (b) Cutting Triangles}\hfill
    \parbox{.15\linewidth}{\centering (c) Snapping}\hfill
    \parbox{.14\linewidth}{\centering (d) Intersection}\hfill
    \parbox{.14\linewidth}{\centering (e) Triangles Subdivision}\hfill
    \parbox{.14\linewidth}{\centering (f) Final Mesh}\hfill
    \parbox{.04\linewidth}{.}\par
    \caption{\revision{Segment insertion into a triangle mesh (a 2D analog of triangle insertion) with and without snapping. (a) Insertion of segment $pq$ into mesh $M$. (b) Identification of cut triangles $\mathcal{T}_I$ (in red). (c) Snapping vertex $v$ to line $pq$ and updating $\mathcal{T}_I$, where $v$ is $\delta$-close to $pq$. $v$ is moved to its closest points on line $pq$ if this does not invert any elements of $M$ (case 1). Vertex $v$ is added in $\mathcal{V}_{\delta}$ both if $v$ is moved (case 1) or not (case 2). The yellow triangle is added to $\mathcal{T}_I$. (d) Computing the intersection of line $pq$ with the edges of $\mathcal{T}_I$ (points $p_1, p_2, p_3$). (e) Triangles requiring subdivision (shown in red). (f) The final mesh after subdivision.}}
    \label{fig:ins_ppl}
\end{figure*}

The key component of our algorithm is three-stage procedure for inserting one triangle \revision{$T$} into a valid tetrahedral mesh $M$, adding new vertices and tetrahedra, and adjusting mesh connectivity, to minimize the number of insertion failures and number of badly shaped tetrahedra created by insertion.   Our algorithm uses ideas from marching tetrahedra \cite{doi1991efficient} and other tetrahedralization methods \cite{George:2003,Weatherill:1994}, as well as marching cubes \cite{Lorensen:1987}. It consists of the following steps: 
\revision{
(1) Find the set $\mathcal{T}_I$ consisting of the tetrahedra of $M$ that triangle $T$ cuts, as defined below; 
(2) Compute the intersection points of the plane spanned by $T$ (denoted as $P$) and the edges of the tetrahedra in $\mathcal{T}_I$;
(3) Subdivide all cut tetrahedra using a connectivity pattern from a pre-computed \emph{tet-subdivision table}. These patterns guarantee that a valid tetrahedral mesh connectivity is maintained.
}

\begin{figure}
    \centering\footnotesize
    \includegraphics[width=\linewidth]{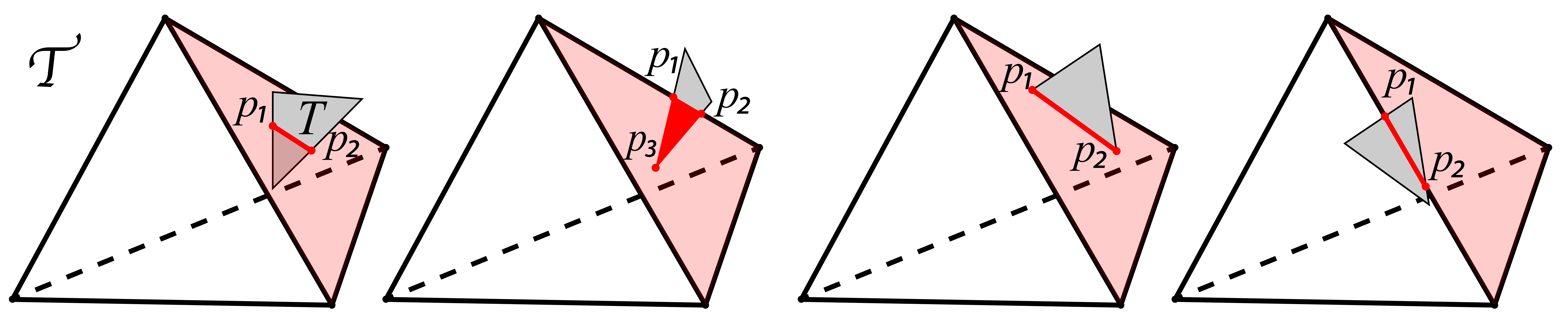}\\
    \parbox{.49\linewidth}{\centering $\mathcal{T}\in{\mathcal{T}_I}$}\hfill
    \parbox{.49\linewidth}{\centering $\mathcal{T}\notin{\mathcal{T}_I}$}\par
    \caption{\revision{Examples of tetrahedra $\mathcal{T}$ included into, or excluded from, $\mathcal{T}_I$. 
    The intersections are shown in red. Left two: $T$ intersects a face of $\mathcal{T}$ at a segment ($[p_1, p_2]$) or a polygon ($[p_1, p_2, p_3]$) that contains interior points of both $T$ and the intersected face of $\mathcal{T}$. In this case, we put $\mathcal{T}$ into ${\mathcal{T}_I}$. Right two: $T$ intersects a face of $\mathcal{T}$ (in light red) at a segment ($[p_1, p_2]$) that does not contain any interior points of either $T$ or the intersected face of $\mathcal{T}$. In this case, we do not put $\mathcal{T}$ into ${\mathcal{T}_I}$.}}
    \label{fig:cut-through}
\end{figure}

\paragraph{\revision{Finding Cut Tetrahedra.}}

\revision{We first define that object $A$ \emph{cuts though} object $B$ if their intersection contains interior points of both $A$ and $B$. We say that triangle $T$ \emph{cuts} tetrahedron $\mathcal{T}$ if (1) it is completely contained inside $\mathcal{T}$, or (2) it cuts through at least one face of $\mathcal{T}$ (Figure~\ref{fig:cut-through}). We initialize $\mathcal{T}_I$ to be the set of the tetrahedra of $M$ that $T$ cuts. Note that this set will be iteratively expanded by the algorithm.}

\revision{
We use exact predicates \cite{shewchuk97a,geogram} for checking if a triangle is contained inside a tetrahedron. To detect if one triangle cuts through another, we combine the exact predicates with the algorithm \cite{guigue:hal-00795042}. The use of predicates ensures topological correctness when using floating-point coordinates.}



\paragraph{\revision{Plane-Tetrahedra Intersection.}}

\begin{figure}
    \centering\footnotesize
    \includegraphics[width=\linewidth]{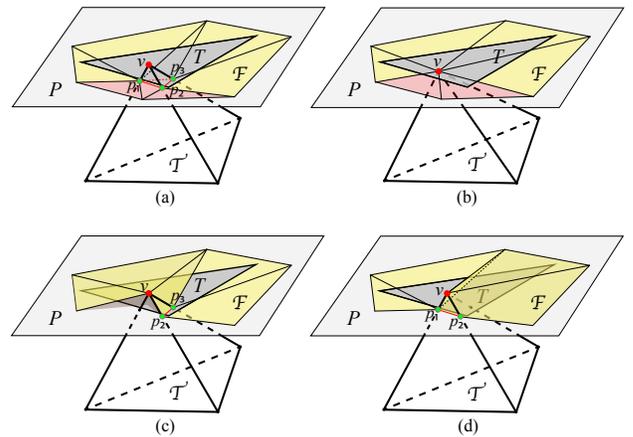}\\
    \caption{\revision{Plane $P$ intersects $\mathcal{T}_I$ ($\mathcal{T}\in\mathcal{T}_I$). (a) The faces of $\mathcal{F}$ (marked in yellow) are the covering of triangle $T$. (b) Snapping $v$ to its closest point on $P$ and expanding $\mathcal{F}$ to include red triangles makes $\mathcal{F}$ safely covering $T$. (c) Snapping a boundary vertex $p_1$ of $\mathcal{F}$ to $v$ changes the area of $\mathcal{F}$ and make it not covering $T$. (d) Snapping an interior vertex $p_3$ of $\mathcal{F}$ to $v$ does not change the boundary of $\mathcal{F}$: $\mathcal{F}$ is still covering $T$.}}
    \label{fig:covering}
\end{figure}

\revision{
To insert a triangle $T$ defining a plane $P$ into $\mathcal{T}_I$ (Figure~\ref{fig:ins_ppl}(c)(d)), we need to ensure that after the insertion: (1) for every point $p\in T$ there is a   face $F$ of the refined tets in the set $\mathcal{T}_I$ such that $\min_{p \in F} \|p-q\|<\delta$; (2) the projection of faces $F$ in $\mathcal{T}_I$, that are within the distance $\delta$ from $T$ to the plane $P$ covers $T$.  We call sets with these properties \emph{covers} of $T$. We allow the cover of triangles to deviate from $P$. This  is crucial for robustly inserting triangles using floating point computations: without it, we observe a significantly higher running time due to more insertion failures, which leads to additional iterations of mesh optimization.  Also, more faces remain uninserted in the final output.
For the first pass of triangle insertion (i.e., before any mesh improvement is done), we use a larger $\delta = \max(\epsilon_\mathrm{zero},10^{-3}{\epsilon})$, while for all subsequent passes we reduce it to $\delta = \epsilon_\mathrm{zero}$. }

\revision{
We start with the idealized case of infinite-precision arithmetics. In this case, we can easily realize the covering of $T$ by intersecting all the faces of the tetrahedra in $\mathcal{T}_I$ with plane $P$. This generates a planar polygonal mesh $\mathcal{F}$  on $P$ covering $T$ and the maximal distance from $T$ to $\mathcal{F}$ is zero (Figure~\ref{fig:covering} (a)).  The vertices of $\mathcal{F}$ are intersection points of $P$ and edges of $\mathcal{T}_I$.}

\revision{When the floating point representation is used to represent the coordinates of vertices, round-off error may result in degenerate or inverted tetrahedra. Our approach is to reject insertion in these cases. However, to minimize the number of triangles that have to be rejected, we either snap vertices of the tetrahedral mesh $M$ to $P$ (Figure~\ref{fig:covering} (b)) or snap intersection points to vertices of $M$ (Figure~\ref{fig:covering} (c)(d)).}
\revision{Moving a vertex $v$ of $\mathcal{T}_I$ (thus changing $M$) changes the cover $\mathcal{F}$, because $P$ intersects the edges of $\mathcal{T}_I$ in different locations. As no intersection of $P$ with an edge of $\mathcal{T}_I$ disappears (at most, it may move to an endpoint), and no new intersections appear (other than at the endpoints shared with already intersected edges), the connectivity of $\mathcal{T}_I$ can be viewed as unchanged, possibly with some zero-length edges. We can view snapping vertices of $\mathcal{T}_I$  to $P$ as a deformation of $\mathcal{F}$, keeping it on plane $P$.  
If the affected vertices are in the interior of $\mathcal{F}$, $\mathcal{F}$ still covers $T$ since the boundary of $\mathcal{F}$ does not change. However, if moving $v$ changes the boundary of $\mathcal{F}$, the covering might be invalidated (Figure~\ref{fig:covering} (b)). In this case, \emph{before} moving a boundary vertex, we extend $\mathcal{T}_I$ by adding its 1-ring neighbourhood, intersect it with $P$, and extend $\mathcal{F}$ accordingly. We repeat this process until all affected vertices are in the interior. }

\revision{
Moving the point $v$ to the plane $P$ might not always be possible, since it could invert some tetrahedra in $M$. In these cases, instead of moving $v$ to $P$, we deform $\mathcal{F}$ by moving some vertices of $\mathcal{F}$ to $v$, which is at $\delta$ distance from $P$ by definition (Figure~\ref{fig:covering} (c)(d)). Similarly to the previous case, this operation can only be applied on interior vertices of $\mathcal{F}$. We thus extend $\mathcal{T}_I$ if this operation affects vertices on the boundary of $\mathcal{F}$.
}

\begin{figure}
    \centering\footnotesize
    \includegraphics[width=0.95\linewidth]{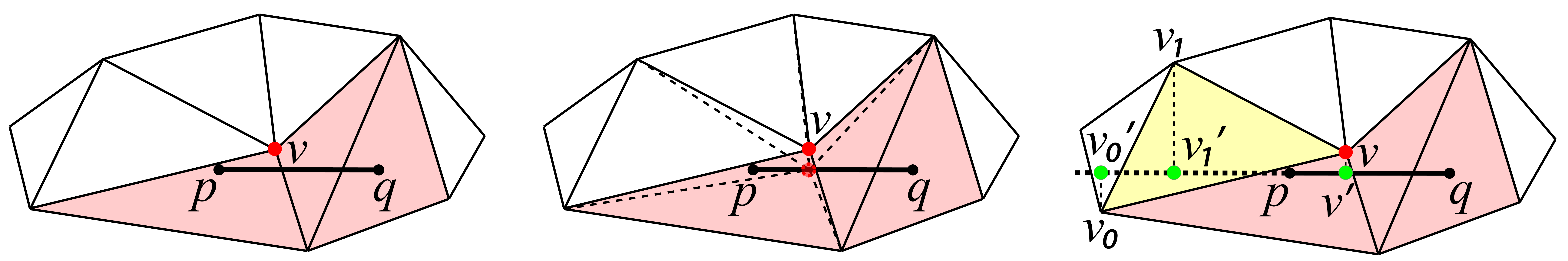}\\
    \parbox{.34\linewidth}{\centering (1)}\hfill
    \parbox{.3\linewidth}{\centering (2)}\hfill
    \parbox{.34\linewidth}{\centering (3)}\par
    \caption{\revision{2D illustration of step (1) to (3) of snapping when inserting segment $[p, q]$. The cut triangles in $\mathcal{T}_I$ are marked in red. (1) Vertex $v$ is within $\delta$ distance to segment $[p, q]$. (2) Check the effect of moving $v$ to line $[p, q]$. (Vertex $v$ cannot be moved to $[p, q]$ in this case, because a triangle reverts its orientation.)
    (3) One-ring triangle $[v, v_0, v_1]$ (marked in yellow) of $v$ is intersecting with line $[p,q]$, and the projection of its edges $[v, v_0], [v, v_1]$ on line $[p, q]$, segment $[v', v_0'], [v', v_1']$, intersect with segment $[p,q]$.}}
    \label{fig:snapping}
\end{figure}

\revision{
In practice, we never explicitly compute $\mathcal{F}$ on the plane $P$ since it is uniquely defined by the intersection points (Appendix~\ref{app:aftersnap}), but instead use the following 4 steps, that directly determine the vertices of $\mathcal{F}$ (the faces of $\mathcal{F}$ are obtained by table-based refinement of $\mathcal{T}_I$).
\begin{enumerate}
    \item Find all vertices of the tetrahedra in $\mathcal{T}_I$, with distance to $P$ smaller than $\delta$  and put them in $\mathcal{V}_{\delta}$ (e.g., vertex $v$ in Figure~\ref{fig:snapping}(1)).
    \item Move vertices in $\mathcal{V}_{\delta}$ to their closest points on $P$ if it does not invert any elements of $M$ (Figure~\ref{fig:snapping}(2)).
    \item For each vertex in $\mathcal{V}_{\delta}$, add all of its vertex-adjacent tetrahedra to $\mathcal{T}_I$ if these are cut  by $P$ and have faces covering $T$ (i.e., the projection of the face to $P$ intersects with $T$). For example,  $[v, v_0, v_1]$ in Figure~\ref{fig:snapping}(3) is added.
    \item Repeat steps (1) to (3) until no more new tetrahedra are added to $\mathcal{T}_I$.
\end{enumerate}
}


\paragraph{\revision{Table-based Tetrahedron Subdivision.}}
\label{sec:table}

\begin{table}
\caption{\revision{A subset of the tet-subdivision table, the complete table is provided in the additional material. The first row corresponds to the case of  a tetrahedron without cut edges, the second and third to the case of one cut edge, $e_0$ and $e_1$ respectively, and the last row to two cut edges $e_0$ and $e_1$. All tetrahedra shown in the table have same edge label.}}\vspace{-5pt}
\centering
 \begin{tabular}{c|c|c|c} 
\diagbox[innerwidth=2cm]{\textbf{I}}{\textbf{II}} & 0 & 1 & $\cdots$ \\ \hline
$0_{(10)} = 000000_{(2)}$ & \raisebox{-.4\height}{\includegraphics[width=5em]{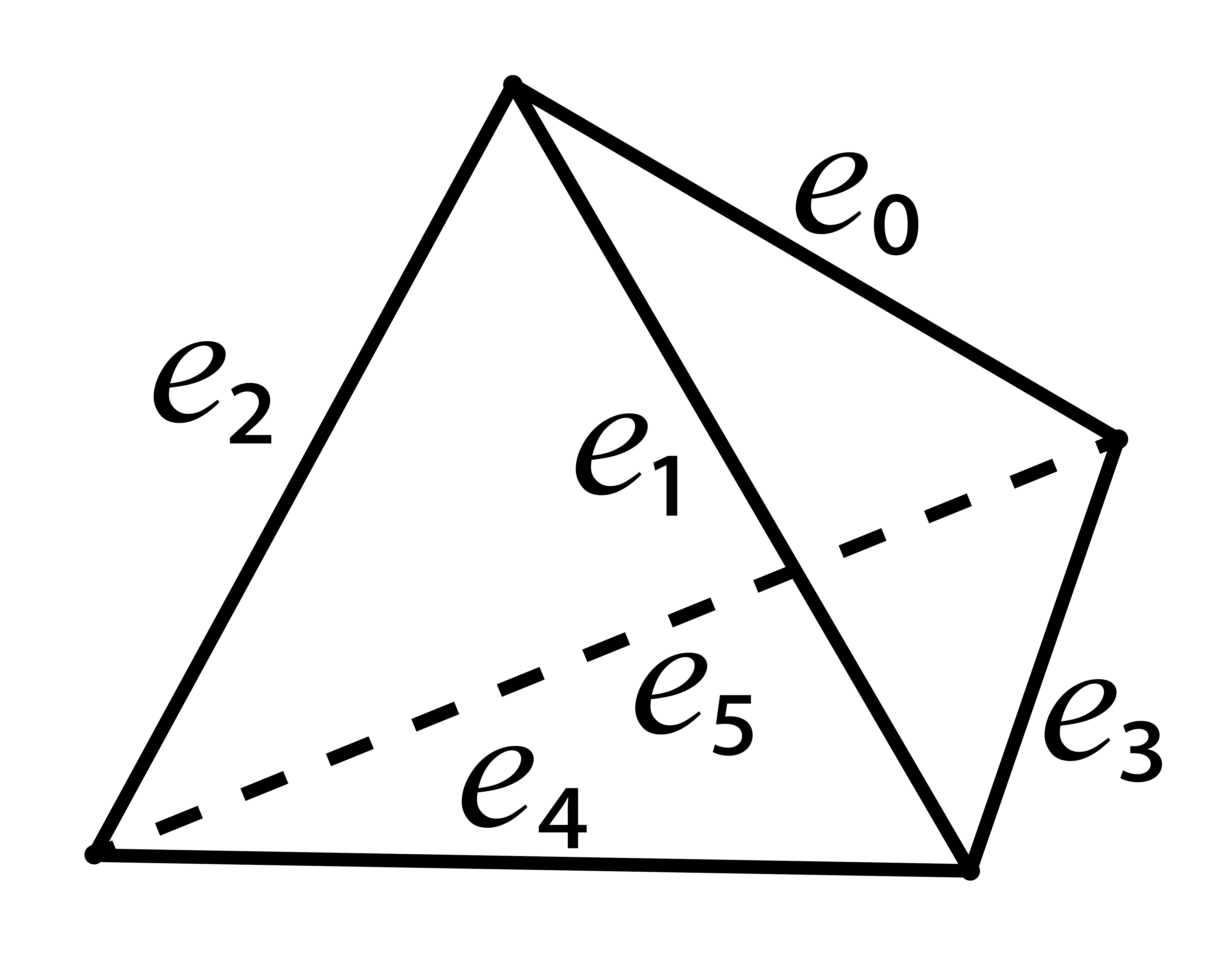}}&  & $\cdots$\\
\hline
$1_{(10)} = 000001_{(2)}$ & \raisebox{-.4\height}{\includegraphics[width=5em]{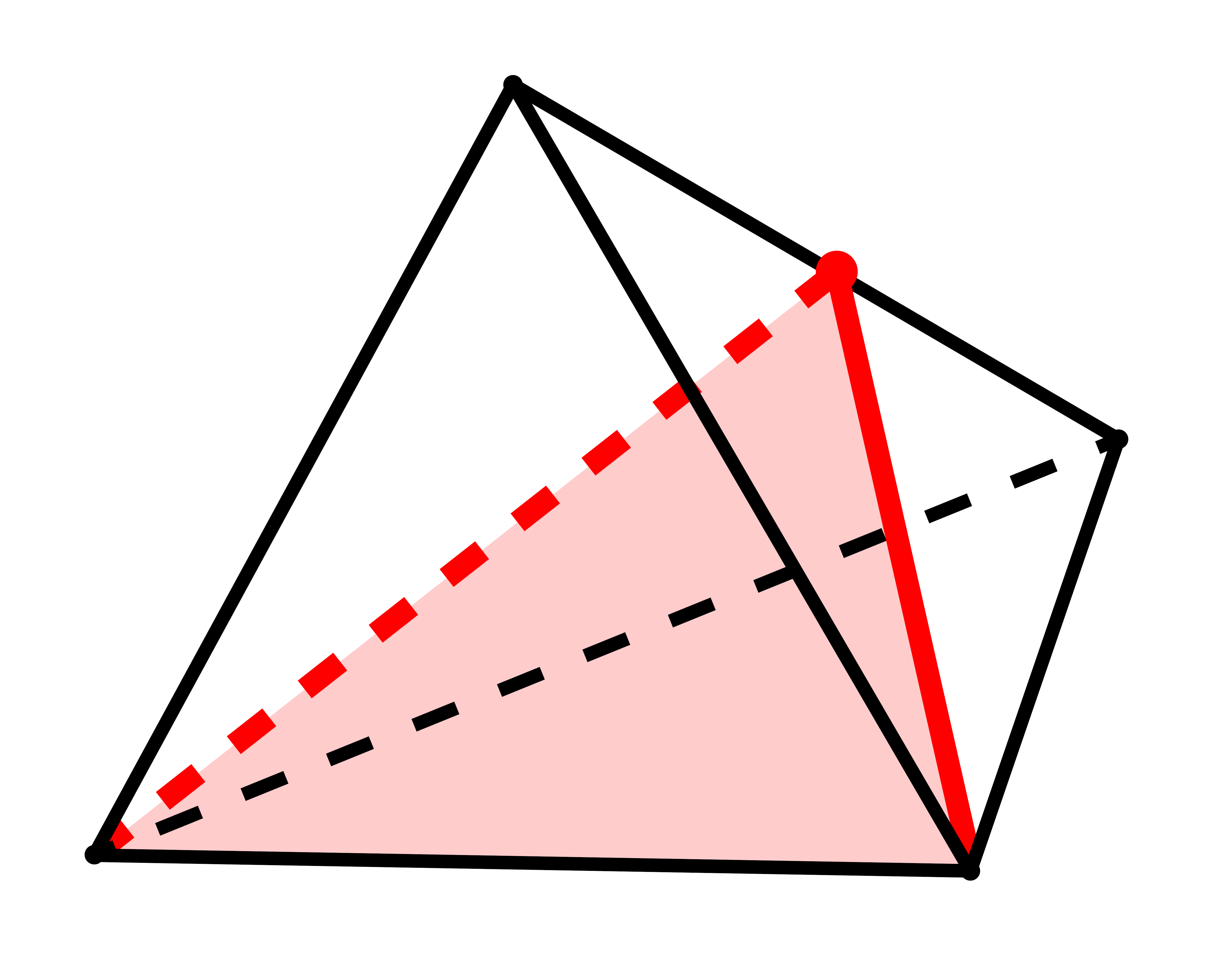}} &  & $\cdots$\\
\hline
$2_{(10)} = 000010_{(2)}$ & \raisebox{-.4\height}{\includegraphics[width=5em]{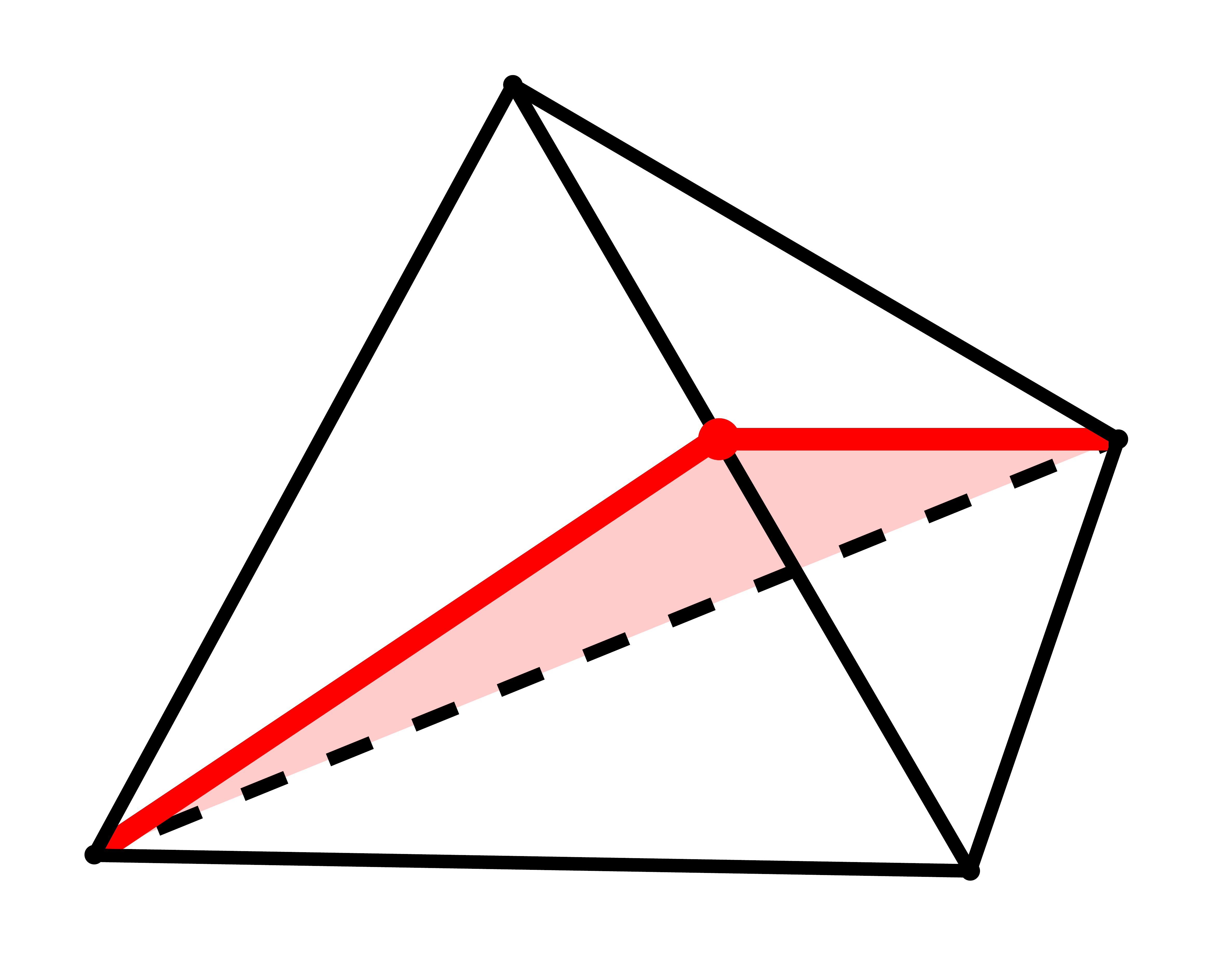}} &  & $\cdots$\\
\hline
$3_{(10)} = 000011_{(2)}$ & 
\raisebox{-.4\height}{\includegraphics[width=5em]{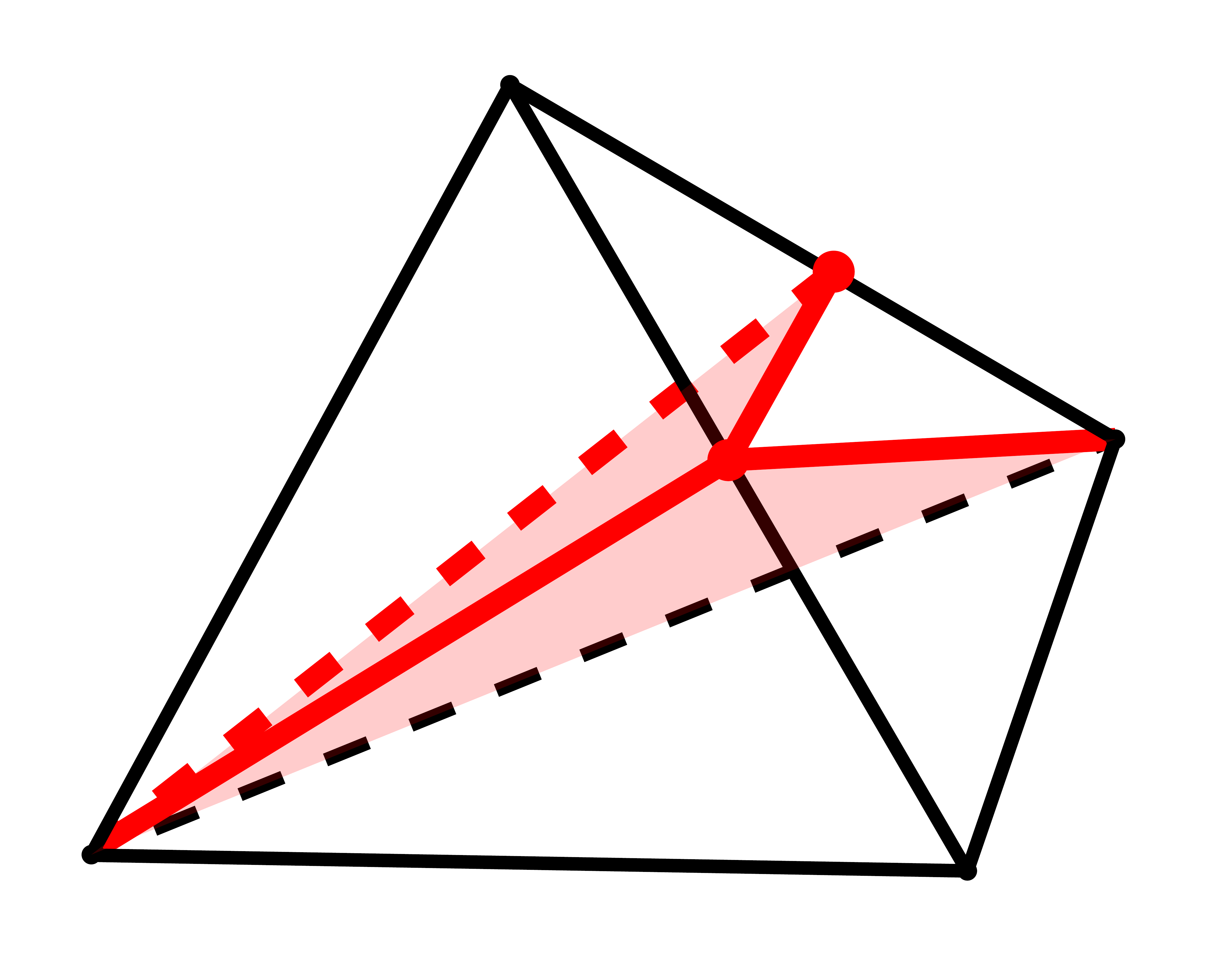}}
& 
\raisebox{-.4\height}{\includegraphics[width=5em]{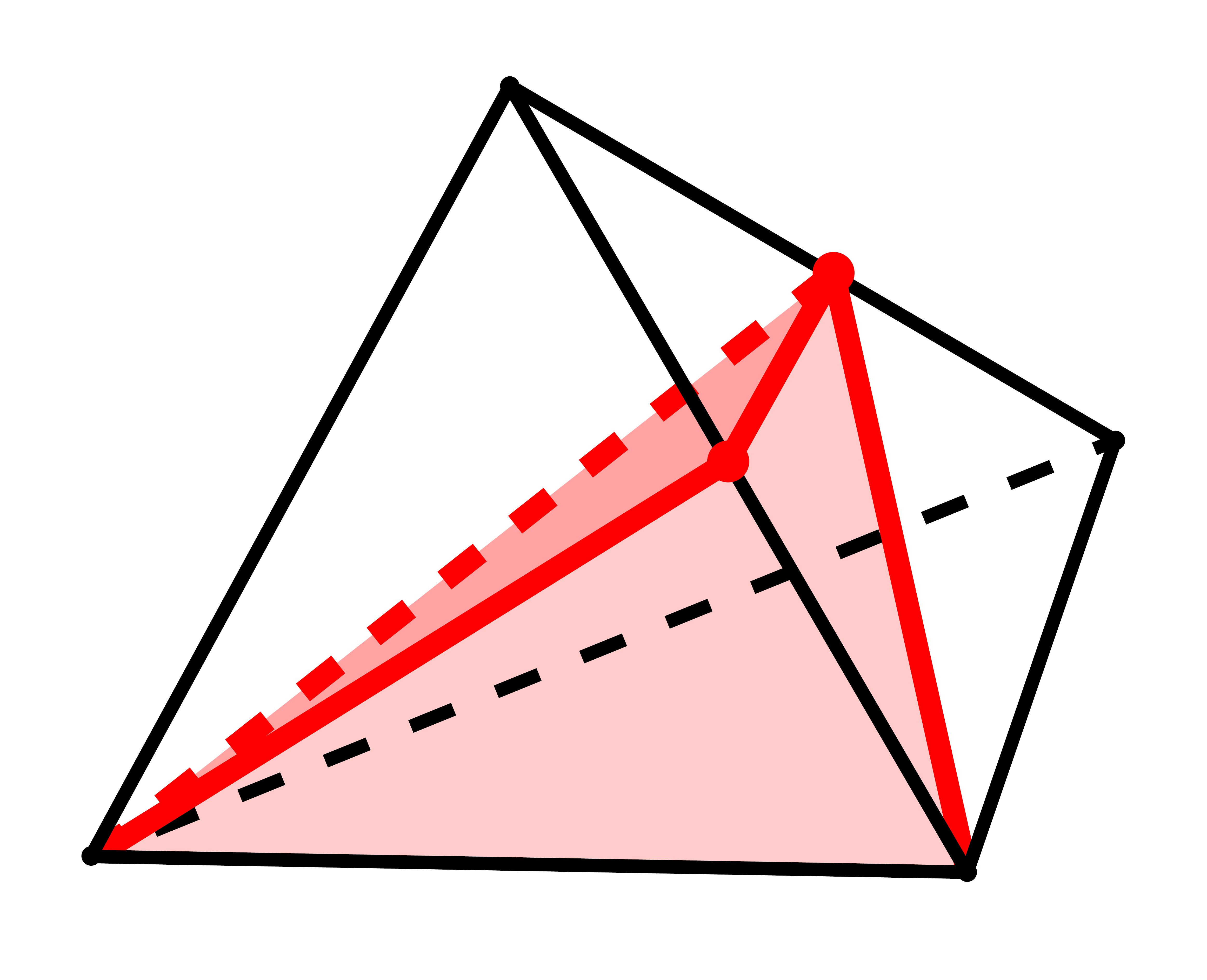}} & $\cdots$\\
\hline
$\vdots$ & $\vdots$ & $\vdots$ & $\ddots$\\\par
\end{tabular}
\label{table:subdivision}
\end{table}

\revision{All tetrahedra sharing the edges of tetrahedra in $\mathcal{T}_I$ cut by $P$ are subdivided into sub-tetrahedra according to the \emph{tet-subdivision} table. Note that this set of tetrahedra usually contains some tetrahedra from $\mathcal{T}_I$ (red elements in Figure~\ref{fig:ins_ppl}(e)) and some neighboring tetrahedra of $\mathcal{T}_I$ (yellow elements in Figure~\ref{fig:ins_ppl}(e)).}

\revision{Since an edge can have at most one intersection point with $P$, the decomposition of the subdivided tetrahedron is largely (but not entirely) determined by which edges are cut. (An edge will be cut if it intersects with $P$ and neither endpoints are snapped.) We record all possible decompositions of a tetrahedron in a subdivision table, indexed by \emph{primary cut indices} and \emph{secondary cut indices}. The primary index (I) (Table \ref{table:subdivision}), is a binary string, indicating which edges are cut. If two edges on a face are cut (3 edges of a face cannot be cut at the same time), there are two possible triangulations of this face and also multiple decompositions of the tetrahedron; the secondary index (II) is the number of a  specific decomposition (Table \ref{table:subdivision}). A primary index paired with a secondary index retrieves a unique decomposition of a tetrahedron.}

\revision{ For an oriented tetrahedron $\mathcal{T}$, there are $2^6=64$ combinations of possible intersection points on its edges, but not all 64 combinations can be practically realized. A direct enumeration shows that the following edge-cut configurations are impossible: (1) $\mathcal{T}$ has six cut edges, (2) $\mathcal{T}$ has five cut edges, (3) $\mathcal{T}$ has 4 cut edges, and 3 of them are on the same face, and (4) $\mathcal{T}$ has 3 cut edges on the same face. In total, there are ${6\choose 6}+{6\choose 5}+3\cdot 4+4 = 23$ impossible edge-cut configurations.}

\begin{figure*}
    \centering\footnotesize
    \includegraphics[width=0.93\linewidth]{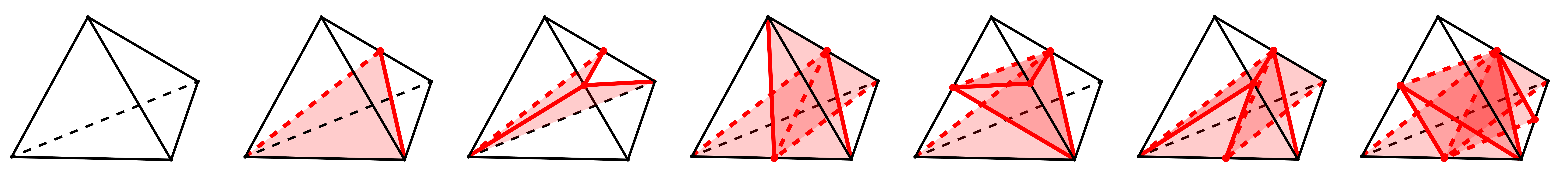}\par
    \parbox{.18\linewidth}{\centering (1) No cut}\hfill
    \parbox{.12\linewidth}{\centering (2) 1 edge cut}\hfill
    \parbox{.13\linewidth}{\centering (3) 2 edge cut}\hfill
    \parbox{.13\linewidth}{\centering (4) 2 edge cut}\hfill
    \parbox{.13\linewidth}{\centering (5) 3 edge cut}\hfill
    \parbox{.12\linewidth}{\centering (6) 3 edge cut}\hfill
    \parbox{.18\linewidth}{\centering (7) 4 edge cut}\par
    \caption{\revision{7 symmetry classes of edge-cut configurations. (4) and (6) can only happen on neighboring tetrahedra of $\mathcal{T}_I$ with only certain edges cut by $P$.}}
    \label{fig:confs}
\end{figure*}

\revision{The remaining 41 realizable edge-cut configurations cover all subdivision cases and we can categorize them into 7 symmetry classes (Figure~\ref{fig:confs}). Five of them were discussed in~\cite{Schweiger:2016} and used for a tetrahedron cut by a plane. We need 2 extra configurations (Figure~\ref{fig:confs}, (4) and (6)) for subdividing the neighboring tetrahedra of $\mathcal{T}_I$ with only certain edges cut by $P$ (yellow elements in Figure~\ref{fig:ins_ppl}(e)).}

\revision{We retrieve a list of decompositions of $\mathcal{T}$ corresponding to a primary index; we now need to select a secondary index corresponding to a decomposition that preserves the validity of mesh $M$ after the subdivision, that is, we want $M$ to have a valid topology and no inverted tetrahedra. To ensure a valid topology, two adjacent }
\begin{wrapfigure}{r}{.25\columnwidth}
    \begin{center}
    \vspace{-\intextsep}
    \hspace*{-.75\columnsep}\includegraphics[width=.25\columnwidth]{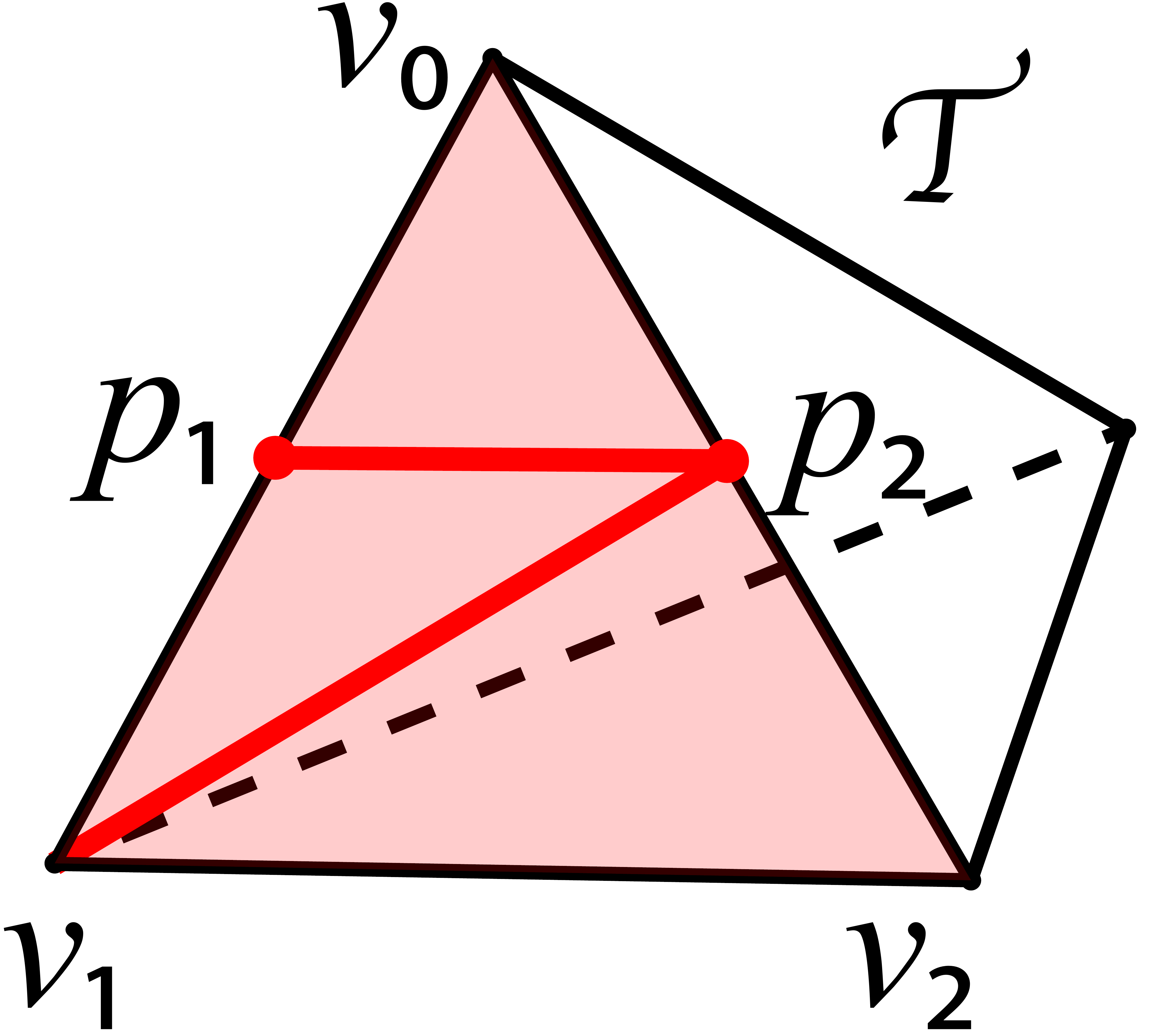}
    \vspace{-\intextsep}
    \end{center}
\end{wrapfigure}
\revision{subdivided tetrahedra must have the same triangulation on the shared face. We set a rule for choosing such triangulation using only the global ordering of the vertices of $M$.}
\revision{For a face $[v_0, v_1, v_2]$ of tetrahedra $\mathcal{T}$ with two intersection points $p_1, p_2$ on it (see inset), we select the triangulation containing the edge $[p_2,v_1]$ if the unique integer label of vertex $v_1$ is larger than the one of $v_2$.}
\revision{Otherwise, we select the triangulation containing the edge $[p_1, v_2]$. This simple rule completely identifies a secondary index and preserves the topology of the mesh. For completeness, we remark that some configurations might require additional vertices (Appendix~\ref{app:secondary}). However, our rule automatically excludes them. We attach the visualization of the tet-subdivision table in the supplementary material.}
\revision{We then check if all sub-tetrahedra have volume larger than $\epsilon_\mathrm{zero}^3$ (since we observed that elements with positive but extremely small volume could delay later insertions in this local region) and reject the insertion if this is not the case.
}

\subsubsection{\revision{Open-boundary Edge Preservation.}}


\revision{After triangle insertion, the input edges shared by two non-coplanar triangles are preserved through the insertion of adjacent triangles, as the plane of the next inserted triangle will intersect the cover $\mathcal{F}$ of a previously inserted triangle. 
This does not hold for boundary edges. An edge is an \emph{open-boundary edge} if it has only one incident triangle or has multiple coplanar incident triangles on the same side of the edge in their common plane. 
} 


\revision{
To preserve an open-boundary edge $e$ of a triangle $T$, we project $e$ and the cover $\mathcal{F}$ of $T$ to the plane $P'$ that best fits the faces of $\mathcal{F}$. Then, we compute the intersection of the projection of $e$ to $P'$ with the projections of the faces of $\mathcal{F}$.
The intersection points of the projection of $e$ and $\mathcal{T}$ are then computed on $P'$ and are lifted back to the corresponding faces of $\mathcal{F}$. Since we have a set of edges cut into two, we can subdivide the affected tetrahedra using the previous table-based tetrahedron subdivision. An example can be found in Appendix~\ref{app:open}.}

Note that the open-boundary edge preservation might fail due to numerical reasons, in this case we rollback the operation and postpone the insertion of the open-boundary triangle to later stages.



\begin{figure}
    \centering\footnotesize
    \includegraphics[width=\linewidth]{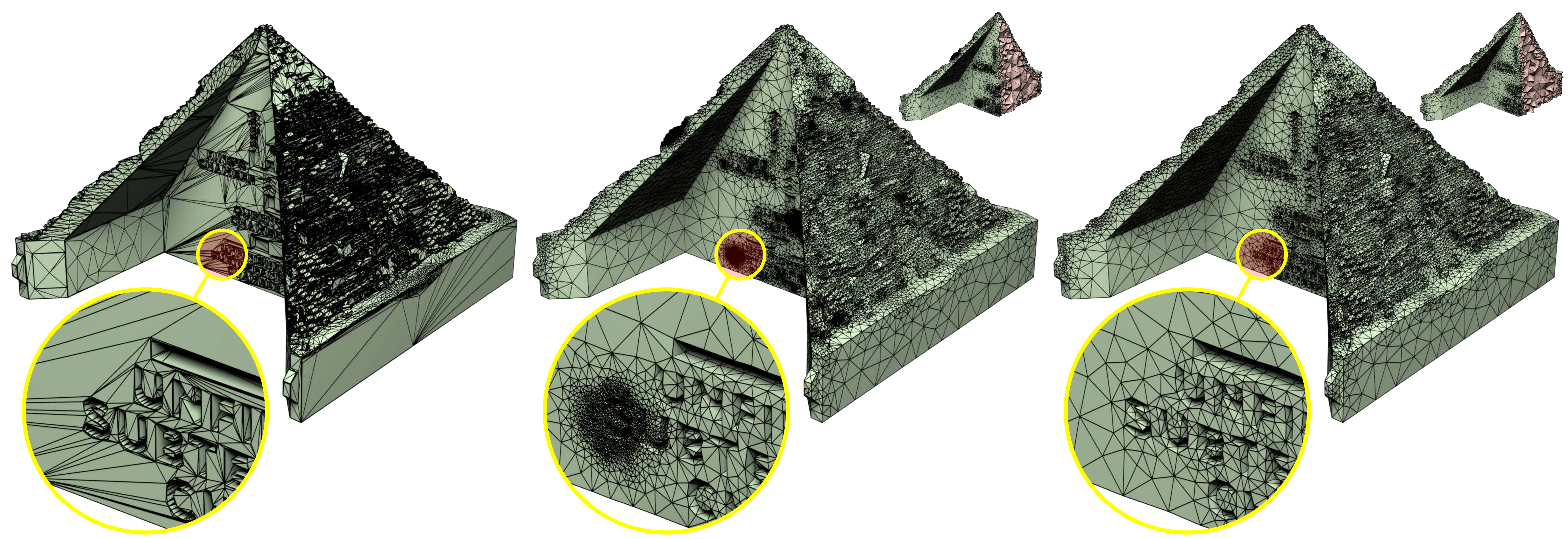}\\
    \parbox{.33\linewidth}{\centering Input\\\#F = 212\,748}\hfill
    \parbox{.33\linewidth}{\centering Unstable energy\\computation, 826s\\\#T = 753\,755\\Max energy = 8.0}\hfill
    \parbox{.33\linewidth}{\centering Stable energy\\computation, 253s\\\#T = 290\,973\\Max energy = 8.0}\par
    \caption{\revision{Example of model where the numerical instability of the AMIPS energy causes over-refinement (middle). By evaluating the energy using rational numbers (when it is above $10^8)$ the issue disappears (right). 
    }}
    \label{fig:instability-overrefinement}
\end{figure}

\subsection{Mesh Improvement}
\label{sec:met:improvement}
We adapt the mesh improvement framework proposed in TetWild~\cite{Hu:2018} that optimizes the conformal AMIPS 3D energy \cite{Rabinovich:2017} to increase the mesh quality, but avoid the overhead introduced by the hybrid kernel by specializing the framework for floating point computation. \revision{Note that, as mentioned in~\cite{Hu:2018}, we use the AMIPS energy since it is differentiable and scale-invariant.}
We made three changes to the original optimization: 
\begin{enumerate}
    \item We try to insert the uninserted input faces every three mesh improvement iterations until all input faces are inserted or the mesh improvement terminates.
    \item We parallelize the vertex smoothing step using a simple graph coloring strategy.
    \item \revision{We discovered an instability in the evaluation of the AMIPS energy computation in floating points, which sometimes leads to overrefinement in TetWild. We propose a fix using a hybrid evaluation that uses rational numbers to compute intermediate results.}
\end{enumerate}

\revision{(1) is a change required by the new algorithm, since not all faces can be inserted when computations are done in floating point. (2) is a minor, yet effective, modification that slightly improves performance (Figure \ref{fig:parallel-comparison}). (3) is a subtle problem, which we now explain in more detail. The conformal AMIPS 3D energy is a Jacobian-based energy defined as: 
\begin{equation}\label{eq:amips}
    \mathrm{AMIPS} = \frac{\mathrm{tr}(\mathbf{J}^T\mathbf{J})}{\mathrm{det}(\mathbf{J})^{2/3}},
\end{equation}
where $\mathbf{J}$ is the Jacobian of the transformation from a regular tetrahedron to the tetrahedron $\mathcal{T}$.
The larger the energy is, the worse the quality of $\mathcal{T}$ is. The minimal value is 3,  the energy of a regular tetrahedron. 
The AMIPS energy is invariant under permutation of the vertices of $\mathcal{T}$, however its numerical evaluation in floating-point arithmetic is not, due to floating-point rounding. Usually the difference is negligible, but when the energy of $\mathcal{T}$ is large (on the order of $10^8$), the floating point computation becomes unstable and the resulting energy could differ by two orders of magnitude, which means that the descent direction that appears to be decreasing the energy may be determined incorrectly (see Appendix~\ref{app:energy} for a concrete example). 
This numerical instability might prevent mesh improvement and thus lead to over-refinement, since the algorithm is trying to add degrees of freedom unnecessary to improve the quality (Figure~\ref{fig:instability-overrefinement}).}

\revision{To address this issue, we raise the energy to the third degree making it completely rational, and evaluate it using rational computation for elements with energy larger than $10^8$. We round the computed value of its third degree to the 64-bit floating point representation, and then compute the cubic root. The rational computation is more accurate (and permutation invariant) but significantly slower. However, the cases of precision loss in the energy are rare, and the overall computational overhead is negligible.
This change has a major effect on the speed and effectiveness of our mesh optimization (Figure~\ref{fig:instability-overrefinement}), avoiding unnecessary refinement and decreasing the overall runtime.}

The mesh improvement terminates when either a user-specified mesh quality or a user-controlled number of iterations is reached. To ensure a fair comparison, for the large dataset testing and all examples in the paper, we use the same stopping criteria (max AMIPS energy is smaller than 10 or the number of optimization iterations reaches 80) and input parameters (envelope size $\epsilon = 10^{-3}d$, targeted edge length $\ell = d/20$, where $d$ is the diagonal's length of the bounding box of the input mesh) as in  \cite{Hu:2018}.

\begin{figure}
    \centering\footnotesize
    \includegraphics[width=\linewidth]{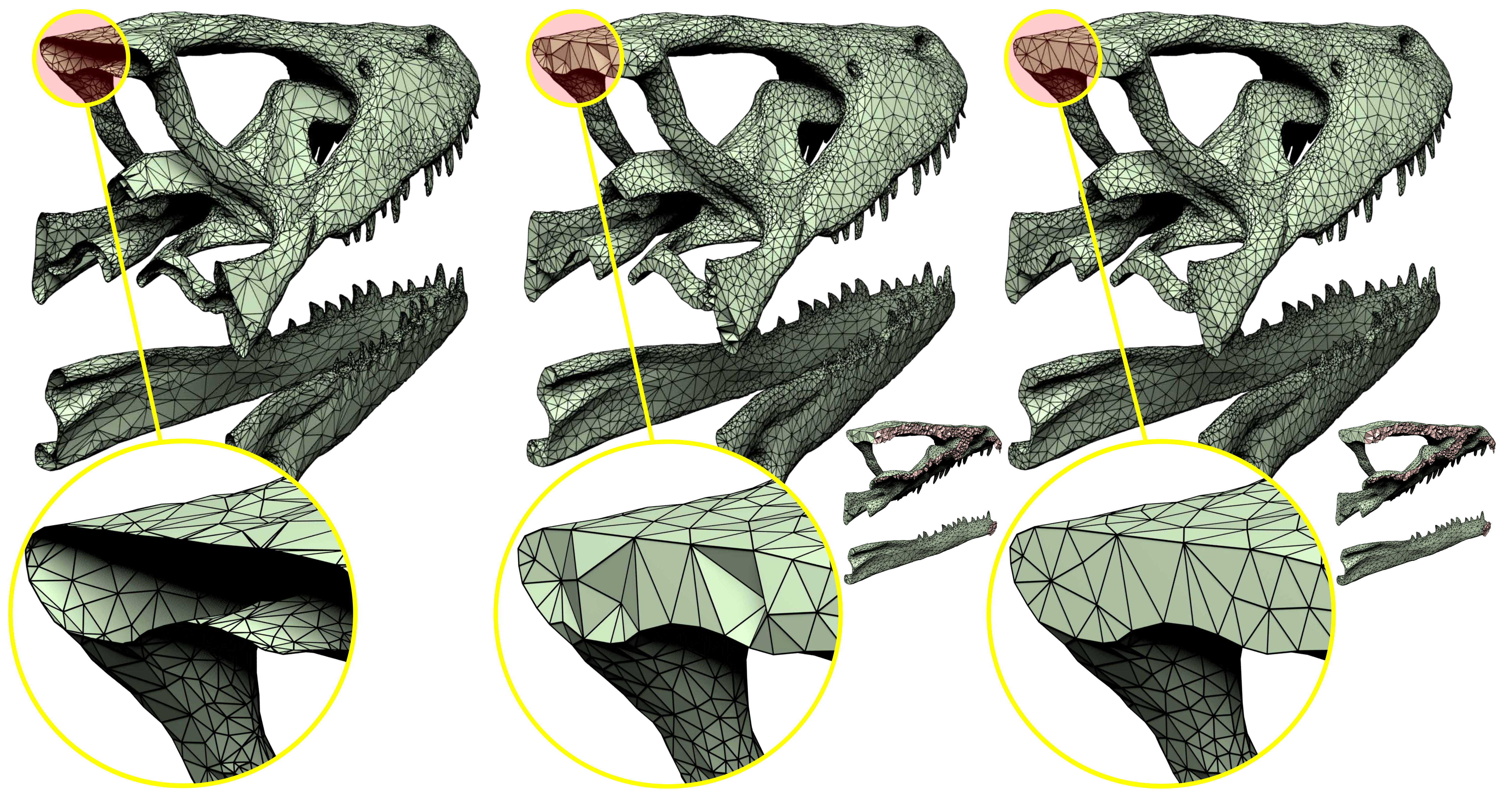}\\[0.4em]
    \parbox{.3\linewidth}{\centering Input\\\#F = 35\,459}\hfill
    \parbox{.33\linewidth}{\centering Output 99s\\\#T = 72\,242\\Max energy = 7.9}\hfill
    \parbox{.33\linewidth}{\centering Output with open region smoothed 113s\\\#T = 72\,094\\Max energy = 10}\par
    \caption{\revision{Input with open boundary on the bottom (left). The output tetrahedral mesh preserves the input geometry and closes the open side (middle). Users can choose to enable an additional smoothing process for smoothing the open region (right).}}
    \label{fig:open}
\end{figure}

\revision{We use the same strategy as in TetWild for handling inputs with open boundaries. We track the vertices on the open boundary and project them back to the open boundary during mesh improvement (Section~\ref{sec:met:improvement}). Elements are classified as inside or outside the surface using the generalized winding number (Section~\ref{sec:met:filtering}).
An example of input with open boundary is shown in Figure~\ref{fig:open}.}

\subsection{Filtering}
\label{sec:met:filtering}
 The output of the mesh improvement step is a volumetric tetrahedral mesh of the expanded bounding box of the input triangle soup, with the preprocessed input triangles inserted. We provide two ways of optionally filter the result, targeting two different applications.
%
The first strategy uses the \revision{fast} winding number \cite{Barill:2018} to filter the tetrahedra outside of the preserved/tracked input \cite{Hu:2018}. %
The second application is a volumetric extension of the mesh arrangement algorithm \cite{Zhou:2016}. In this case, the input becomes a set of triangle soups, coupled with a set of Boolean operations to perform on them. During the triangle insertion stage, we keep track of the provenance of each triangle, and use it at the end to compute a set of generalized winding numbers (one for each tracked input surface) for the centroids of all tetrahedra in the volumetric mesh. 
We use the set of winding numbers 
to decide which tetrahedron to keep by checking if it is supposed to be contained in the result of the Boolean operation.
For instance, when intersecting two triangle soups, we keep all tetrahedra that are inside both input triangle soups, according to the winding number definition. 
 
\revision{There are three major advantages of this method over \cite{Zhou:2016}: (1) Boolean operations can be performed on non-PWN surfaces, (2) the output is equipped with a tetrahedral mesh, which could be useful in downstream applications, and (3) the surface quality is high since the algorithm is allowed to remesh within the $\epsilon$ envelope.}

\begin{figure*}
    \centering\footnotesize
    \includegraphics[width=\linewidth]{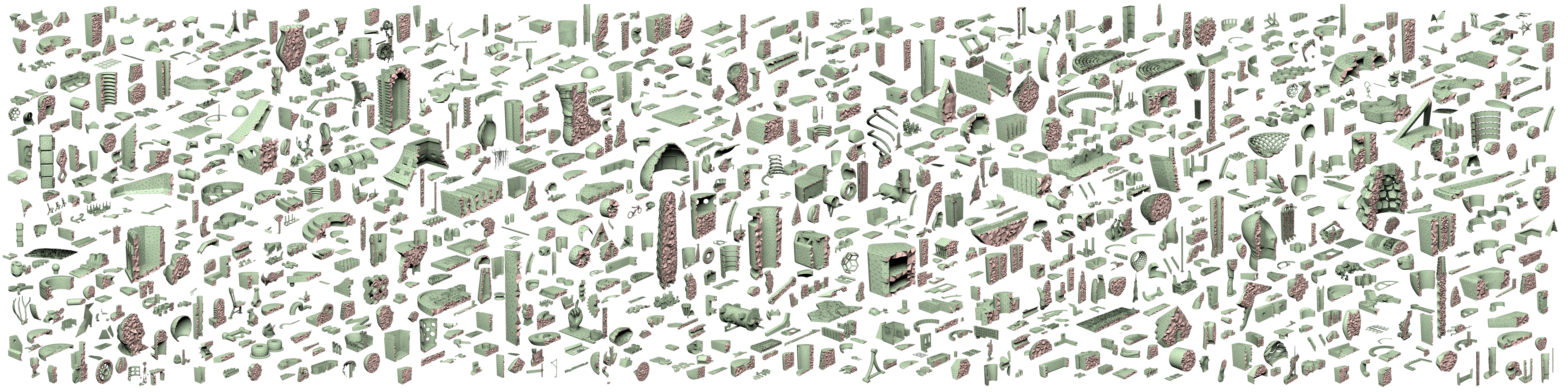}\par
    \caption{\revision{1000 random samples of \textsc{fTetWild} output on Thingi10k dataset.}}
    \label{fig:1k}
\end{figure*}

\begin{figure}
    \centering\footnotesize
    Memory usage of \textsc{fTetWild} (MB)
    \includegraphics[width=0.9\linewidth]{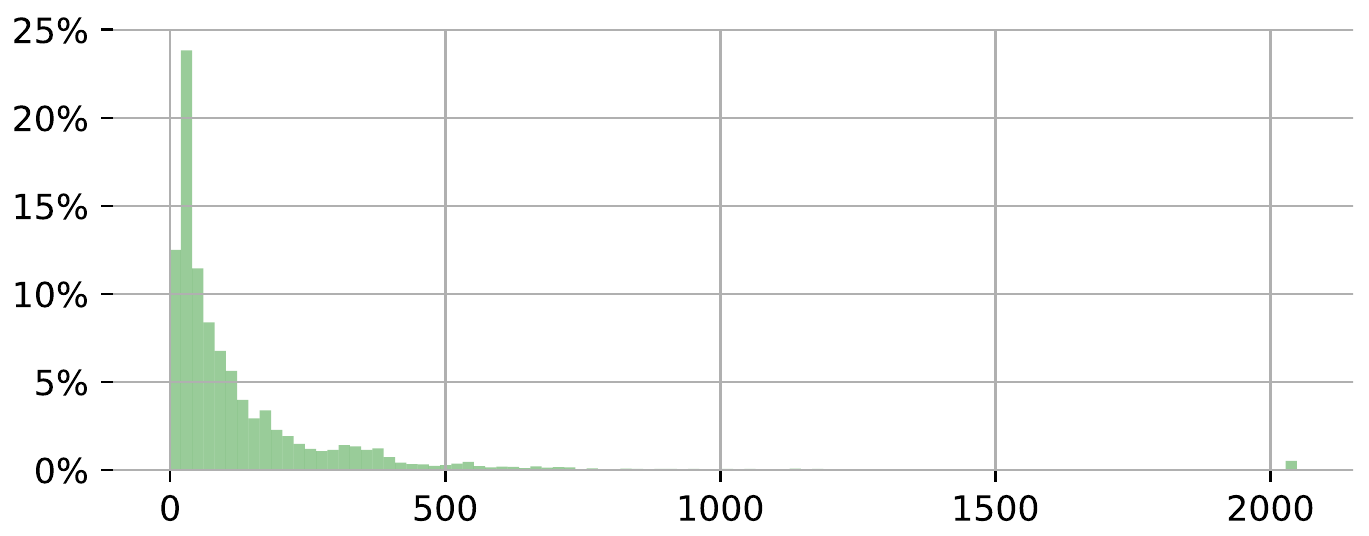}\par
    \caption{Histogram of memory usage of \textsc{fTetWild} over the Thingi10k dataset \revision{(data truncated at 2GB)}.}
    \label{fig:memory}
\end{figure}

\section{Results}
\label{sec:results}
Our algorithm is implemented in C++ and uses Eigen \cite{eigenweb} for the linear algebra routines. We perform a large-scale comparison of our method with other meshing methods on the Thingi10k dataset~\cite{Zhou:2016:TKA}, which contains 10\,000 real-world surface triangle meshes. We run our experiments on cluster nodes with a Xeon E5-2690v4 2.6GHz, allowing every model to use up to 8 threads, 128GB memory, and 24 hours running time. The reference implementation and the scripts used to generate the results are attached to the submission and will be released as an open-source project.



\begin{table}
    \caption{\revision{Comparison of code robustness and performance on the Thingi10k dataset.}}
    \centering\small
    \begin{tabular}{c r r r r r r}
    Method & \makecell{Success\\rate}& \makecell{Out of\\memory\\($>$32GB)} & \makecell{Time\\exceeded\\($>$3h)} & \makecell{Algorithm\\limitation}& \makecell{Average\\time(s)}\\
    \hline
        \textbf{CGAL} & 79.00\% & 0.00\% & 0.00\% & 21.00\% & 11.7 \\
        \textbf{TetGen} & 49.50\% & 0.10\% & 1.70\% & 48.70\% & 32.3 \\
        \textbf{TetWild} & *99.89\% & 0.05\% & 0.11\% & 0.00\% & 360.0 \\
        \textbf{Ours} & \textbf{**99.97\%} & \textbf{0.02\%} & \textbf{0.03\%} & \textbf{0.00\%} & \textbf{49.8} \\
    \end{tabular}\\
    \footnotesize\emph{Note:} The maximum resources allowed for each model are 3 hours and 32GB of memory. The first 3 lines of data are from~\cite{Hu:2018}, Table 2. \revision{Note that the average time (last column) is computed over all the models for which each method succeeded, and it is thus not directly comparable between different methods.} *: TetWild exceeds the 3h time on 11 models. If 27 hours of maximal running time are allowed, TetWild achieves 100\% success rate. \revision{**: Our method exceeds the 3h time limit on 3 models. If 11 hours of maximal running time are allowed, \textsc{fTetWild} achieves 100\% success rate. }
    \label{tab:comparison}
\end{table}

\subsection{Success Rate}
With the above memory and time constraints, \textsc{fTetWild} successfully tetrahedralizes \revision{100\%} of the 10\,000 input meshes (Figure~\ref{fig:1k}). Most of the input models can be tetrahedralized with less than 1GB of RAM as detailed in Figure~\ref{fig:memory}. Note that very complex models might require more memory, for instance the one in Figure~\ref{fig:ntop} uses around 17GB of memory. 

\revision{As observed in~\cite{Hu:2018}, most of the state-of-the-art tet-meshers have low success rate on \emph{in-the-wild} data. We summarize the results on the whole Thingi10k dataset in Table~\ref{tab:comparison}. Note that only
our method and TetWild have high success rates: our average timing is however seven times faster than TetWild.}

\subsection{Running Time}

\begin{figure}
    \centering
    \footnotesize
    \includegraphics[width=\linewidth, left]{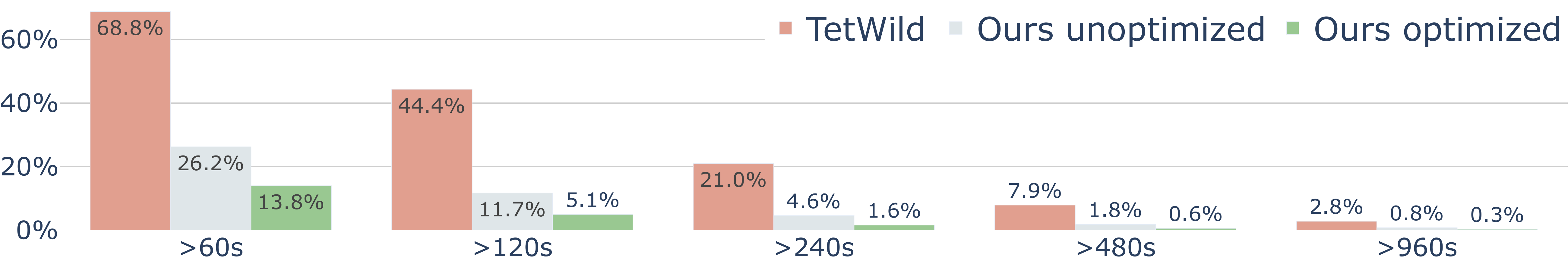}
    \caption{Percentage of models requiring more than a certain time for our parallel and serial algorithm compared with TetWild on Thingi10k dataset.}
    \label{fig:parallel-comparison}
\end{figure}

\begin{figure}
    \centering\footnotesize
    \includegraphics[width=\linewidth]{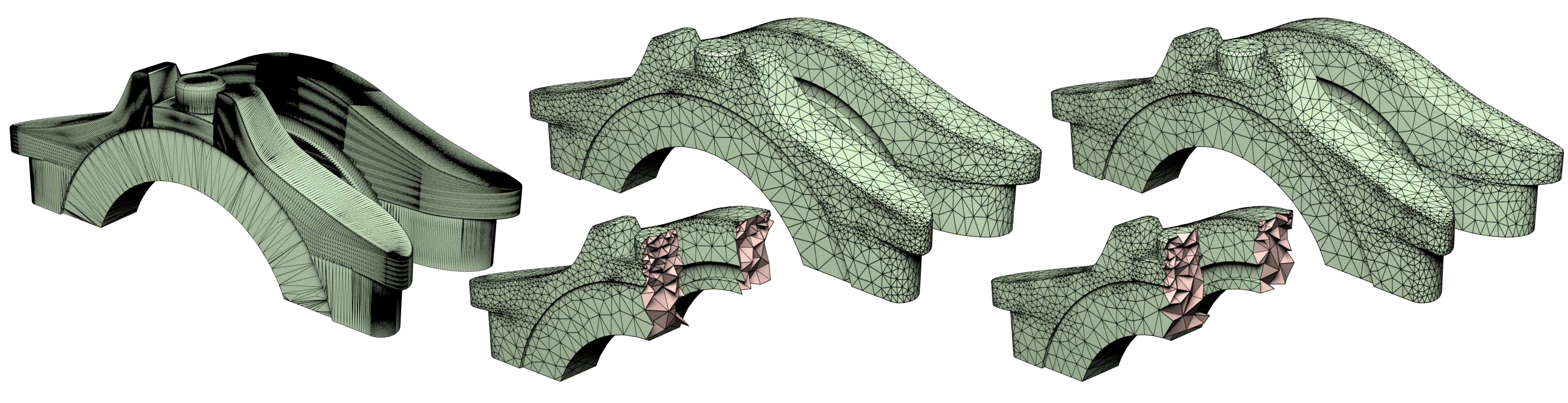}
    \parbox{.3\linewidth}{\centering Input\\\#F = 171\,436}\hfill
    \parbox{.3\linewidth}{\centering TetWild 17hr\\\#T = 39\,312\\Max energy = 1625.4}\hfill
    \parbox{.3\linewidth}{\centering Ours 56m\\\#T = 36\,605\\Max energy = 8.5}\par
    \caption{Example of a challenging model where \textsc{fTetWild} is 17 times faster than TetWild.}
    \label{fig:superfast}
\end{figure}

\paragraph{Thingi10k Dataset (10000 Models)}
\revision{We compare the running time of our method with TetWild. For a fair comparison, we disable our code optimizations that could be easily ported to TetWild, such as parallelization of the preprocessing and smoothing step, and using the recent fast winding number algorithm for the final filtering. Without these optimizations, our algorithm is 4 times faster than TetWild on average (80.4s vs 360s). With code optimizations, we further improve our running time to 49.8s on average on a machine with 8 cores, which is 7 times faster than the serial implementation of TetWild (Figure~\ref{fig:parallel-comparison}). On more complex examples, like the model in Figure~\ref{fig:superfast}, our method is up to 17 times faster than TetWild. }

\paragraph{Reduced Thingi10k Dataset (4540 Models)}
We use a reduced dataset containing the intersection of the Thingi10k models that TetGen, CGAL, TetWild, and our method all succeed on. The dataset contains 4540 models, and allows us to fairly compare the performance of the different methods. On average, our method is \revision{comparable (18.5s)} to the widely used, Delaunay-based tetrahedral mesher TetGen (22s), and is faster than CGAL (95s) and TetWild (107s), while robustly handling imperfect inputs. Figure~\ref{fig:teaser} shows the number of models requiring more than a given time. For example, within less than 2 minutes, our method successfully tetrahedralizes \revision{98.7}\% of the inputs. It is interesting to note that the tail of the distribution of our method is shorter than both TetGen and CGAL. For instance, there are only \revision{4} models where our method requires more than 16 minutes, differently from TetGen, CGAL, and TetWild which have 20, 122, and 25 models, respectively.


\subsection{Mesh Quality}

\begin{figure}
    \centering\footnotesize
    \includegraphics[width=\linewidth]{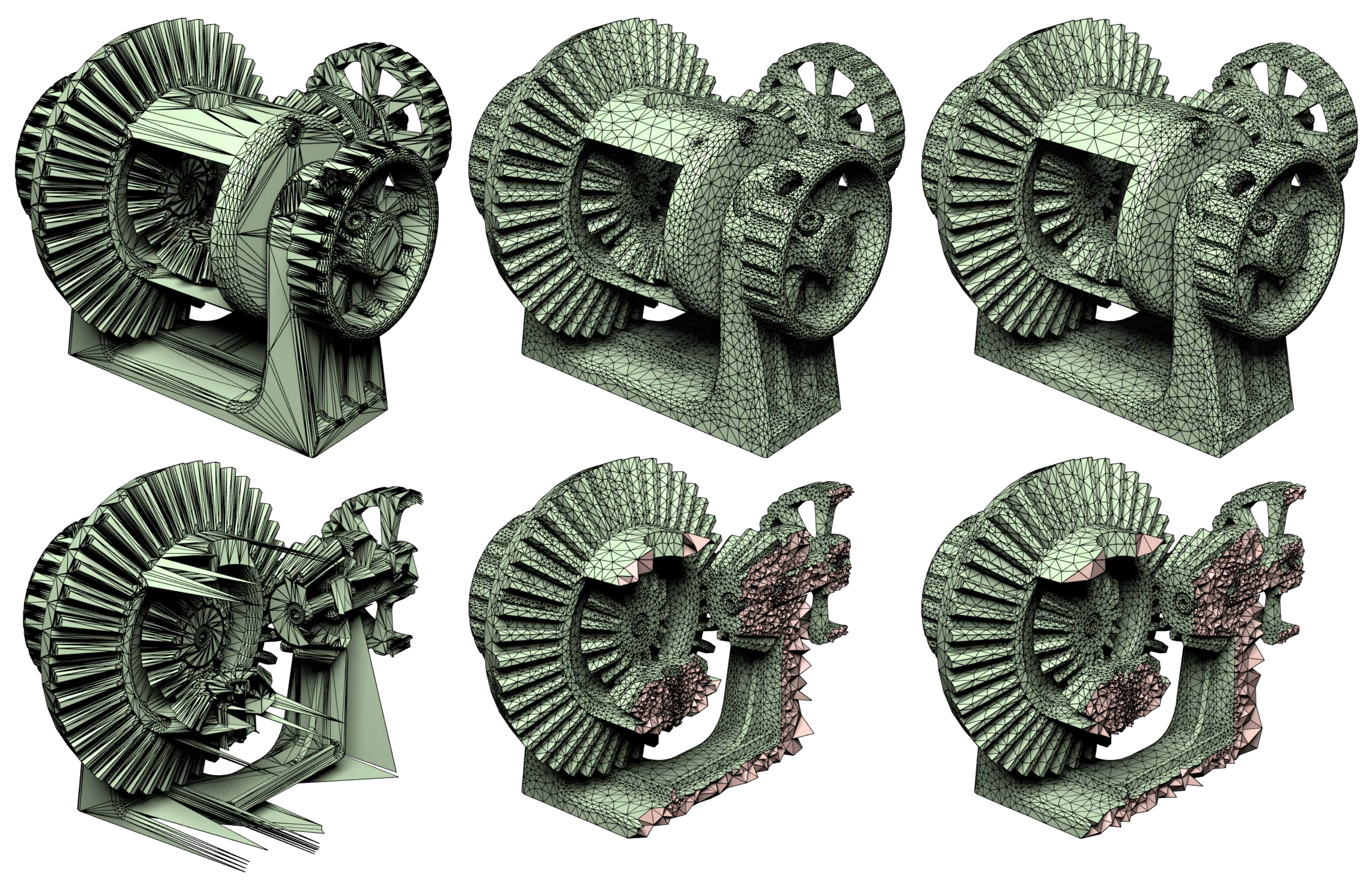}\\[0.5em]
    \parbox{.3\linewidth}{\centering Input\\\#F = 240\,486}\hfill
    \parbox{.3\linewidth}{\centering TetWild 2476s\\\#T = 376\,437\\Max energy = 8.0}\hfill
    \parbox{.3\linewidth}{\centering Ours 411s\\\#T = 311\,318\\Max energy = 8.9}\par
    \caption{Our method (right) produces high-quality tet-meshes that are similar to  TetWild (middle).}
    \label{fig:tetwild}
\end{figure}

\begin{figure}
    \centering\footnotesize
    \parbox{\linewidth}{\centering\footnotesize
    \parbox{.5\linewidth}{\centering TetWild}\hfill
    \parbox{.5\linewidth}{\centering Ours}\par
    \parbox{\linewidth}{\includegraphics[width=\linewidth]{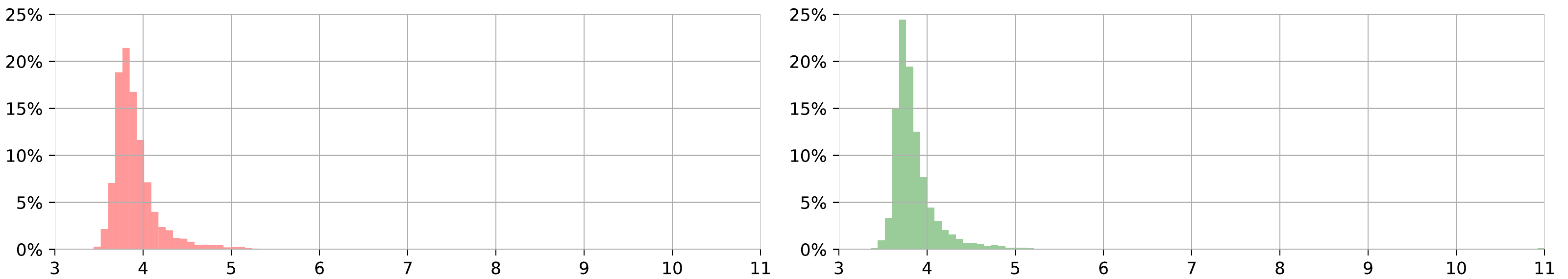}}\\[0.3em]
    Average AMIPS energy\\[0.5em]
    \parbox{\linewidth}{\includegraphics[width=\linewidth]{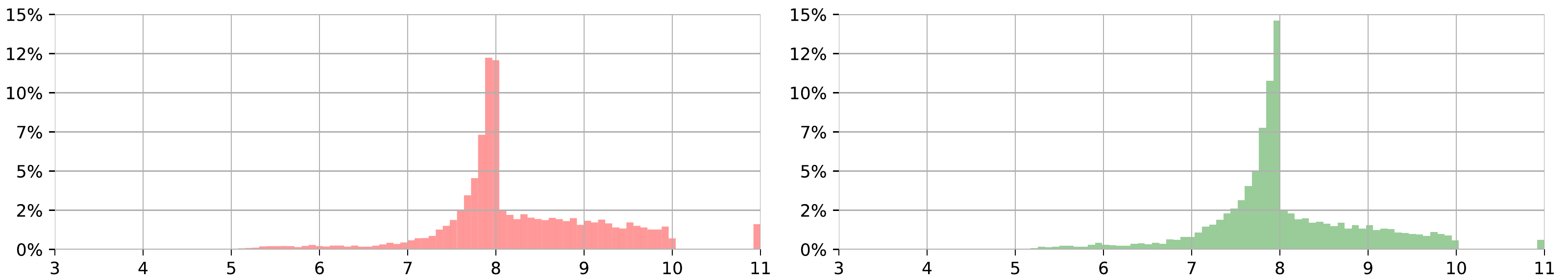}}\\[0.3em]
    Maximum AMIPS energy (truncated at 11)\\[0.5em]
    \parbox{\linewidth}{\includegraphics[width=\linewidth]{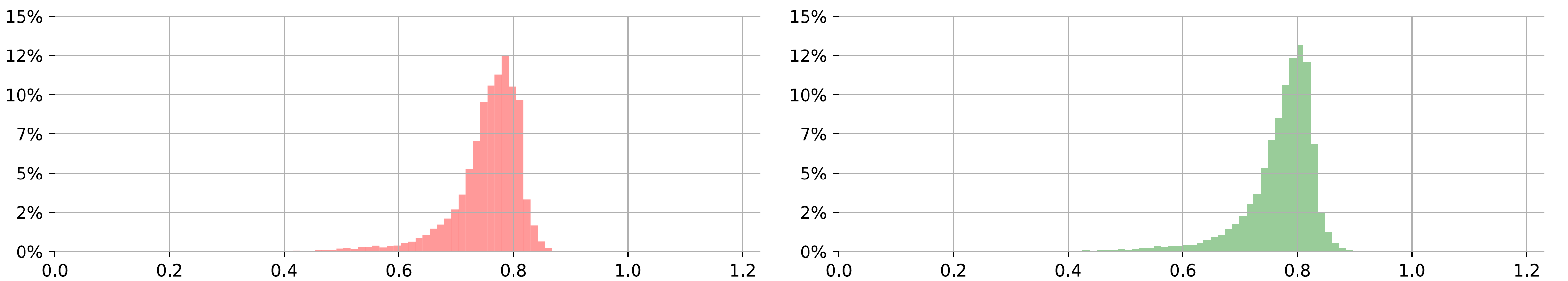}}\\[0.3em]
    Average smallest dihedral angle\\[0.5em]
    \parbox{\linewidth}{\includegraphics[width=\linewidth]{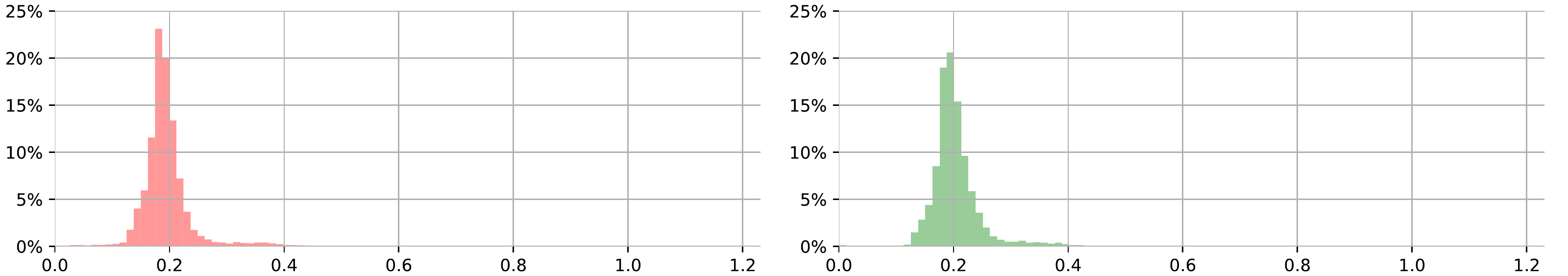}}\\[0.3em]
    Minimum smallest dihedral angle\\[0.5em]
    \parbox{\linewidth}{\includegraphics[width=\linewidth]{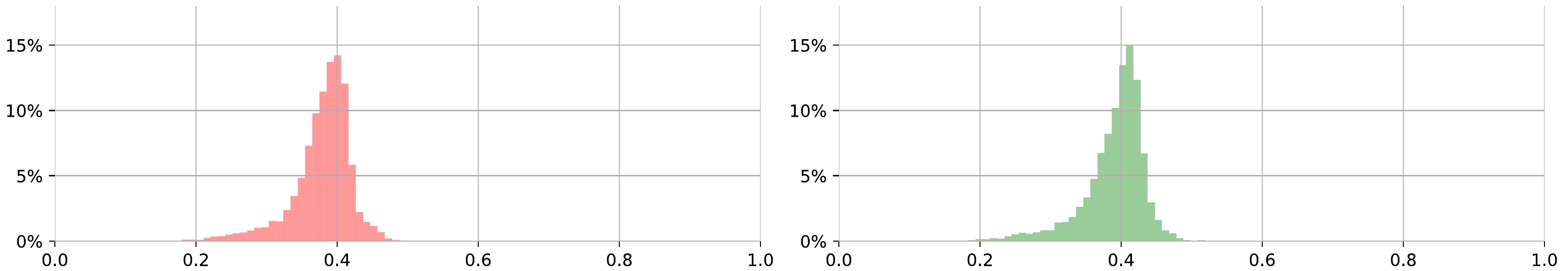}}\\[0.3em]
    Average volume-to-edge ratio\\[0.5em]
    \parbox{\linewidth}{\includegraphics[width=\linewidth]{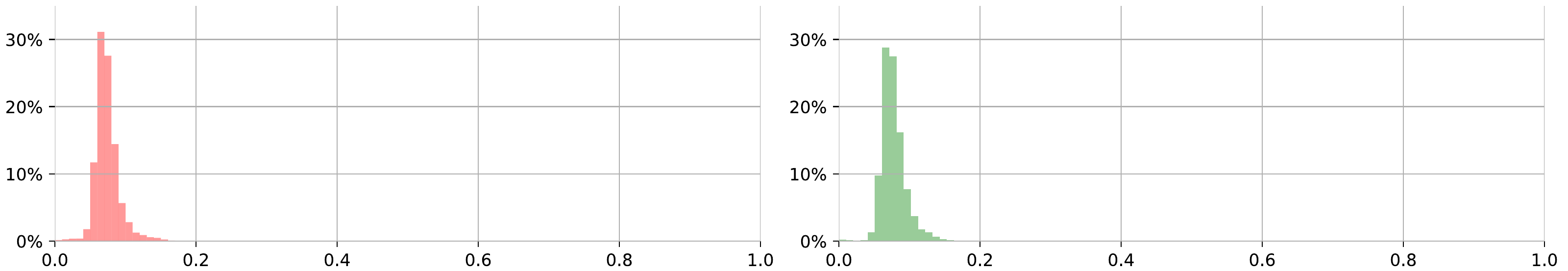}}\\[0.3em]
    Minimum volume-to-edge ratio\\[0.5em]
    \parbox{\linewidth}{\includegraphics[width=\linewidth]{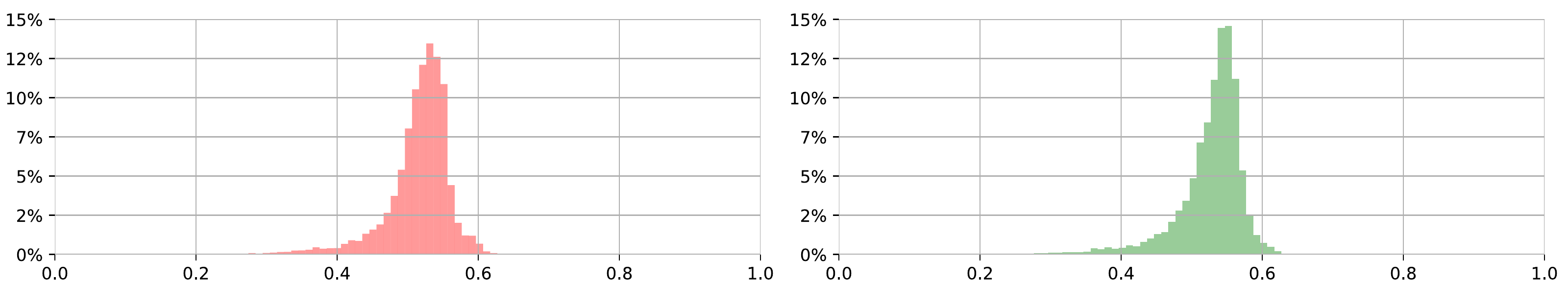}}\\[0.3em]
    Average aspect ratio\\[0.5em]
    \parbox{\linewidth}{\includegraphics[width=\linewidth]{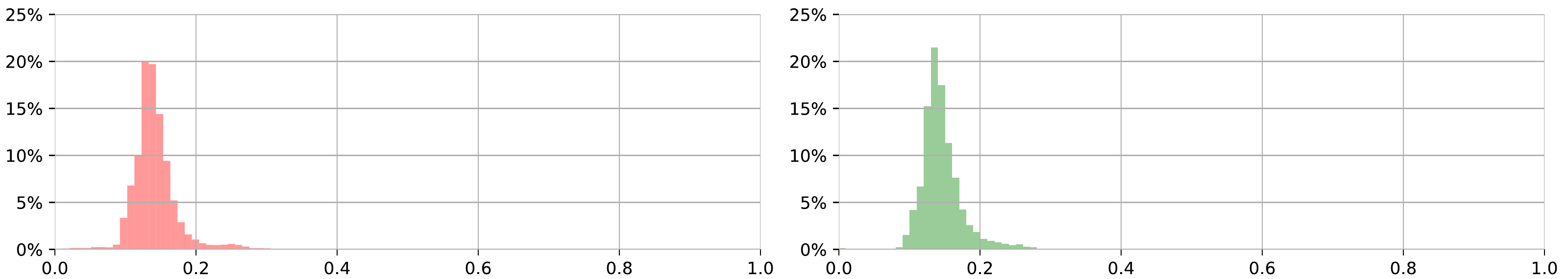}}\\[0.3em]
    Minimum aspect ratio\\[0.5em]
    \parbox{\linewidth}{\includegraphics[width=\linewidth]{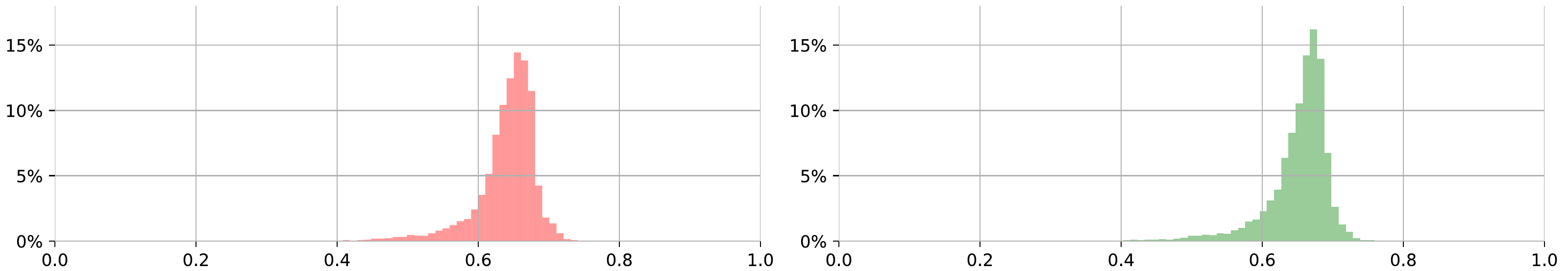}}\\[0.3em]
    Average radius-to-edge ratio\\[0.5em]
    \parbox{\linewidth}{\includegraphics[width=\linewidth]{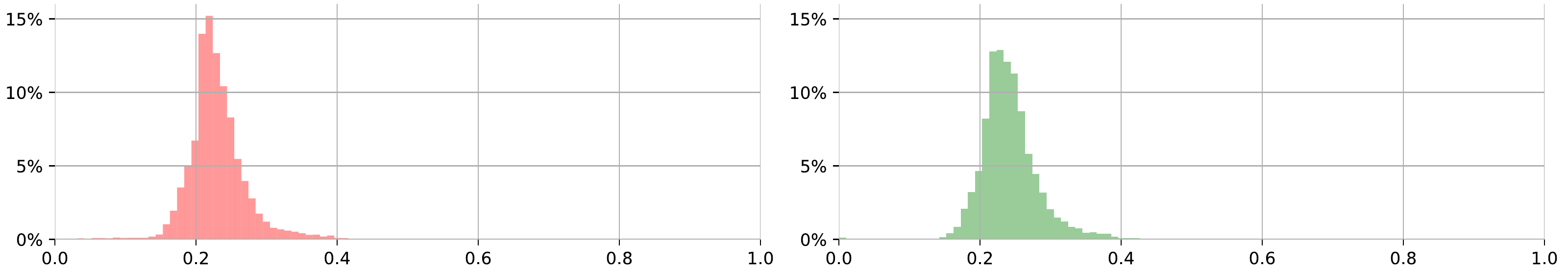}}\\[0.3em]
    Minimum radius-to-edge ratio\\
    }
    \caption{\revision{Histogram for mesh quality comparison of TetWild (red) and our method (green) in five different quality measures. The statistic is based on the output of the whole Thingi10k dataset.}}
    \label{fig:quality}
\end{figure}

\revision{The geometric quality of meshes produced by our algorithm is similar to the meshes produced by TetWild (Figure~\ref{fig:tetwild}), since our method implements a similar mesh optimization strategy. We quantitatively evaluate and compare the element quality of TetWild and our output using five different measures:
\begin{enumerate}
    \item AMIPS energy (Equation~\eqref{eq:amips}), range $[3, +\infty)$, optimal 3,
    \item Minimal dihedral angle, range $(0, 1.23]$, optimal 1.23,
    \item Volume-to-edge ratio $6\sqrt{2}\, V /\ell_{\mathrm{max}}^3$, range $(0, 1]$, optimal 1,
    \item Aspect ratio $\sqrt{3/2}\, h_\mathrm{min}/\ell_\mathrm{max}$, range $(0, 1]$, optimal 1,
    \item Radius-to-edge ratio $2\sqrt{6}\,r_\mathrm{in}/\ell_\mathrm{max}$, range $(0, 1]$, optimal 1,
\end{enumerate}
where $V$ is the volume, $\ell_\mathrm{max}$ is the longest edge, $h_\mathrm{min}$ the minimum height, and $r_\mathrm{in}$ the radius of the inscribed circle of a tetrahedron $\mathcal{T}$. We use (3), (4) and (5) since these are standard measures for tetrahedral quality~\cite{Shewchuk02whatis}.}

\revision{Figure~\ref{fig:quality} shows the histograms of worst and average element quality of 10\,000 output meshes of TetWild and our method. The quality of our outputs are quite similar to TetWild's output. 
We refer to the study in ~\cite[Figure~14]{Hu:2018} for the full quality comparison of TetWild and other tetrahedral meshing algorithms.}

\subsection{Mesh Density}

\begin{figure}
    \centering\footnotesize
    Comparison of mesh size between TetWild and \textsc{fTetWild}\\
    \includegraphics[width=\linewidth]{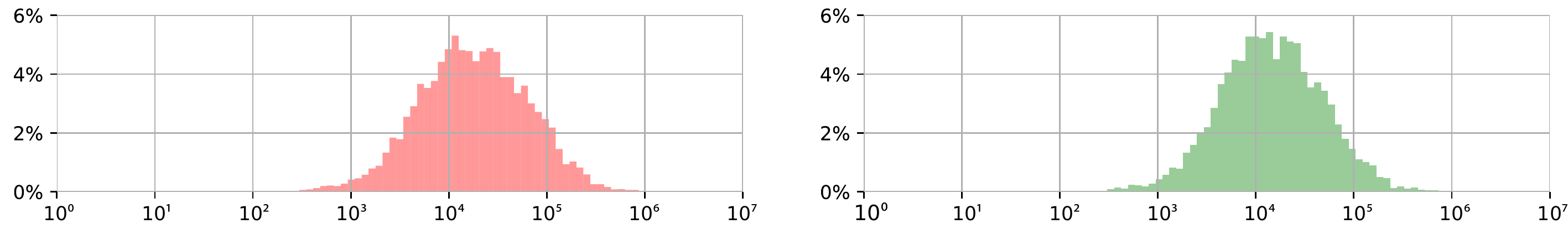}
    \parbox{.45\linewidth}{\centering TetWild}\hfill
    \parbox{.45\linewidth}{\centering Ours}\par
    \caption{\revision{Histograms of number of tetrahedra in log scale for the output meshes of Thingi10k dataset.}}
    \label{fig:mesh_size}
\end{figure}

\revision{Compared with TetWild, our method generates meshes of similar density (Figure~\ref{fig:mesh_size}). Both TetWild and our method aim to generate as-coarse-as possible meshes while preserving the input surface. This choice is useful in downstream applications to reduce their computational cost. Optionally, the algorithm supports a user-specified sizing field to increase the density if desired.}

\revision{In contrast to our method,  TetGen preserves the input surface geometry \emph{exactly} and thus generates a dense tetrahedral mesh around the surface if the input surface mesh is dense, as visible in the model shown in Figure~\ref{fig:teaser}. CGAL approximates the surface by means of an implicit function, but sometimes over-refines  sharp features and tiny artifacts as illustrated in Figure~\ref{fig:teaser}, where the dark spots are over-refined regions.} 

\begin{figure}
    \centering\footnotesize
    \includegraphics[width=\linewidth]{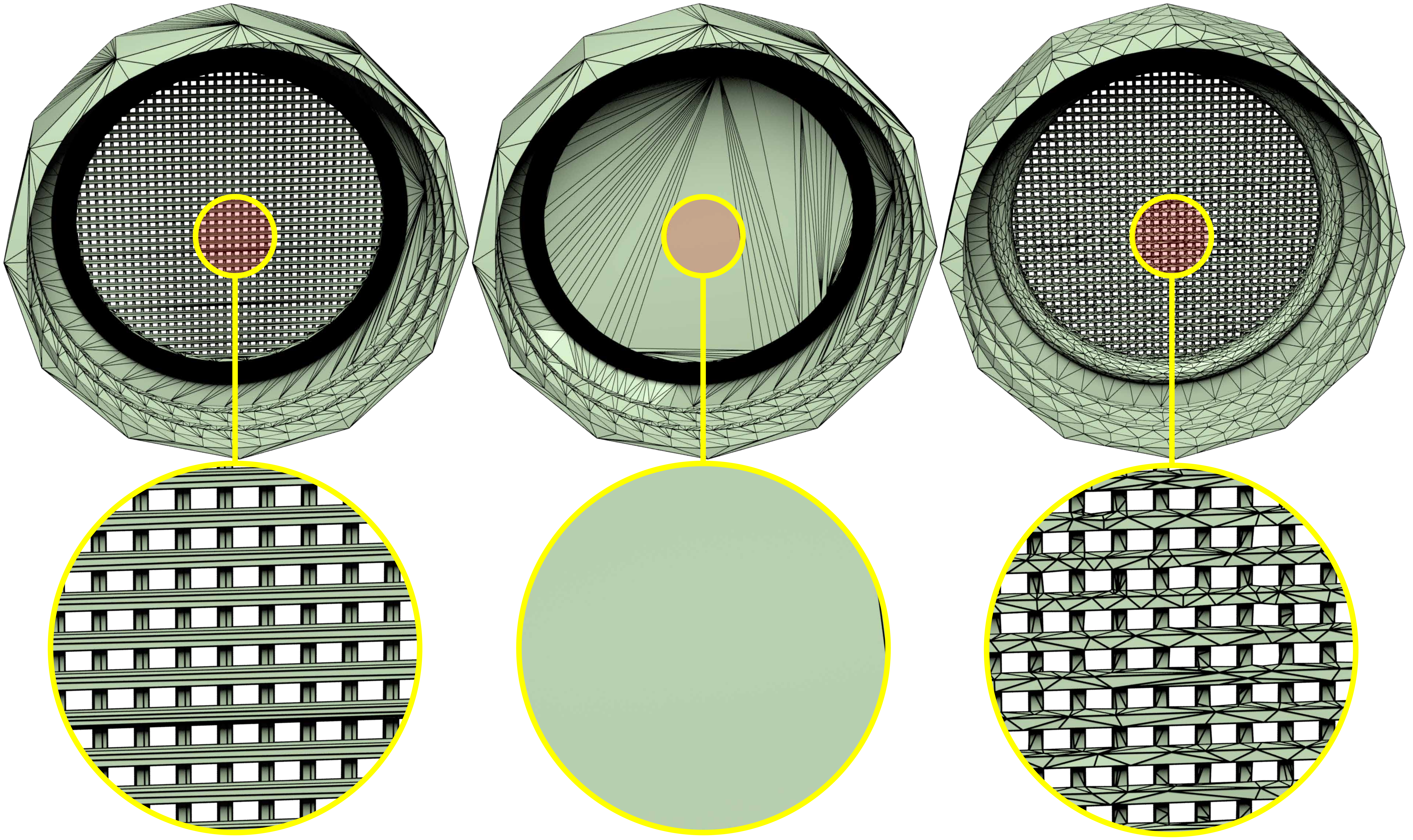}\\[1.em]
    \parbox{.3\linewidth}{\centering Input\\\#F = 16\,248}\hfill
    \parbox{.3\linewidth}{\centering MeshFix 23s\\\#F = 13\,486}\hfill
    \parbox{.3\linewidth}{\centering Our 129s\\\#F = 31\,348}\par
    \caption{Example of repairing an invalid triangular mesh (left) with MeshFix (middle) and our algorithm (right). MeshFix is fast but loses details during processing, while our method preserves them. The max AMIPS energy of our intermediate tetrahedral mesh is 1975. Here we stop mesh improvement when maximum energy reaches 2000. }
    \label{fig:mesh-fix}
\end{figure}

\section{Applications}
\subsection{Mesh Repair}

\begin{figure}
    \centering\footnotesize
    \includegraphics[width=\linewidth]{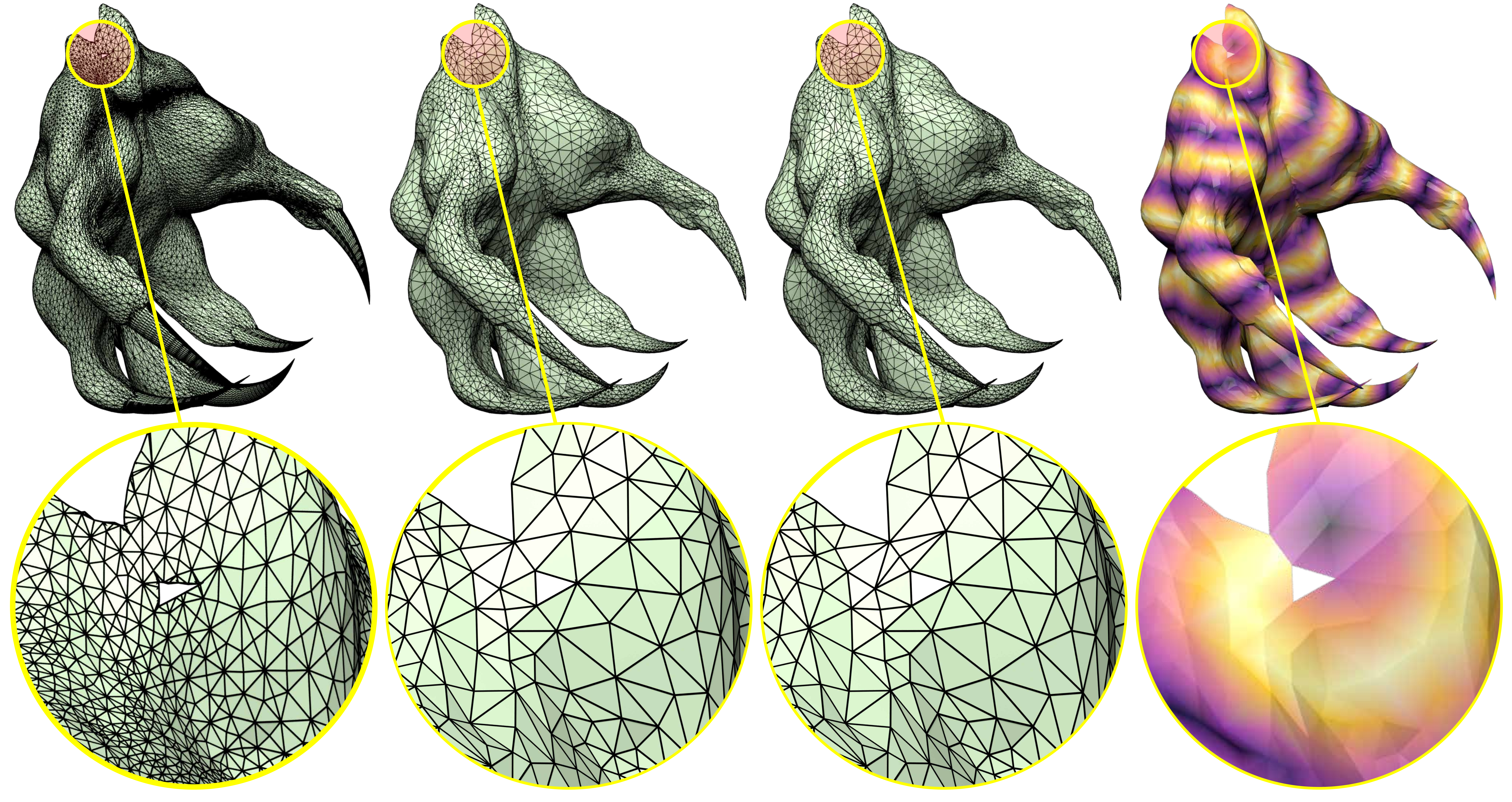}\\[1.em]
    \parbox{.24\linewidth}{\centering Input\\\#F = 16\,248}\hfill
    \parbox{.24\linewidth}{\centering Non-manifold output 87s\\\#F = 42\,924\\Max energy = 9.0}\hfill
    \parbox{.24\linewidth}{\centering Manifold output 90s\\\#F = 42\,931\\Max energy = 9.2}\hfill
    \parbox{.24\linewidth}{\centering Geodesic distance from one point on the manifold surface}\par
    \caption{\revision{Example of a non-manifold surface mesh (left) which is automatically repaired by our algorithm (right second).}}
    \label{fig:manifold}
\end{figure}

Similarly to TetWild, our algorithm can be used to repair imperfect triangle meshes by tetrahedralizing the volume and extracting the surface of the generated tetrahedral mesh. However, the mesh improvement step of our method (Section~\ref{sec:met:improvement}) can be stopped at any time since we  maintain an inversion-free floating point tetrahedral mesh at all  stages of our algorithm. High tetrahedral mesh quality is not required for this application, and we can stop mesh optimization as soon as all input faces are inserted, further reducing the running time. We compared our result with the state-of-the-art mesh repairing tool MeshFix \cite{Attene:meshfix:2010} in Figure~\ref{fig:mesh-fix}. Our method, while slower, provides a higher-quality result with controllable geometric error. \revision{A minor, yet important, observation is that keeping only the boundary of a valid tetrahedral mesh might generate a non-manifold surface mesh (Figure \ref{fig:manifold}). To avoid this problem, we identify the non-manifold edges and split them. Then we duplicate every non-manifold vertex to ensure a global manifold output, using the algorithm proposed in \cite{ATTENE2009850}. Note that this procedure ensures manifoldness, but introduces vertices in the same geometric position. With this minor change, our algorithm can be used to repair triangle meshes, guaranteeing the extraction of an high-quality, manifold boundary surface mesh within the prescribed distance from the input triangle soup.}

\begin{figure}
    \centering\footnotesize
    \includegraphics[width=\linewidth]{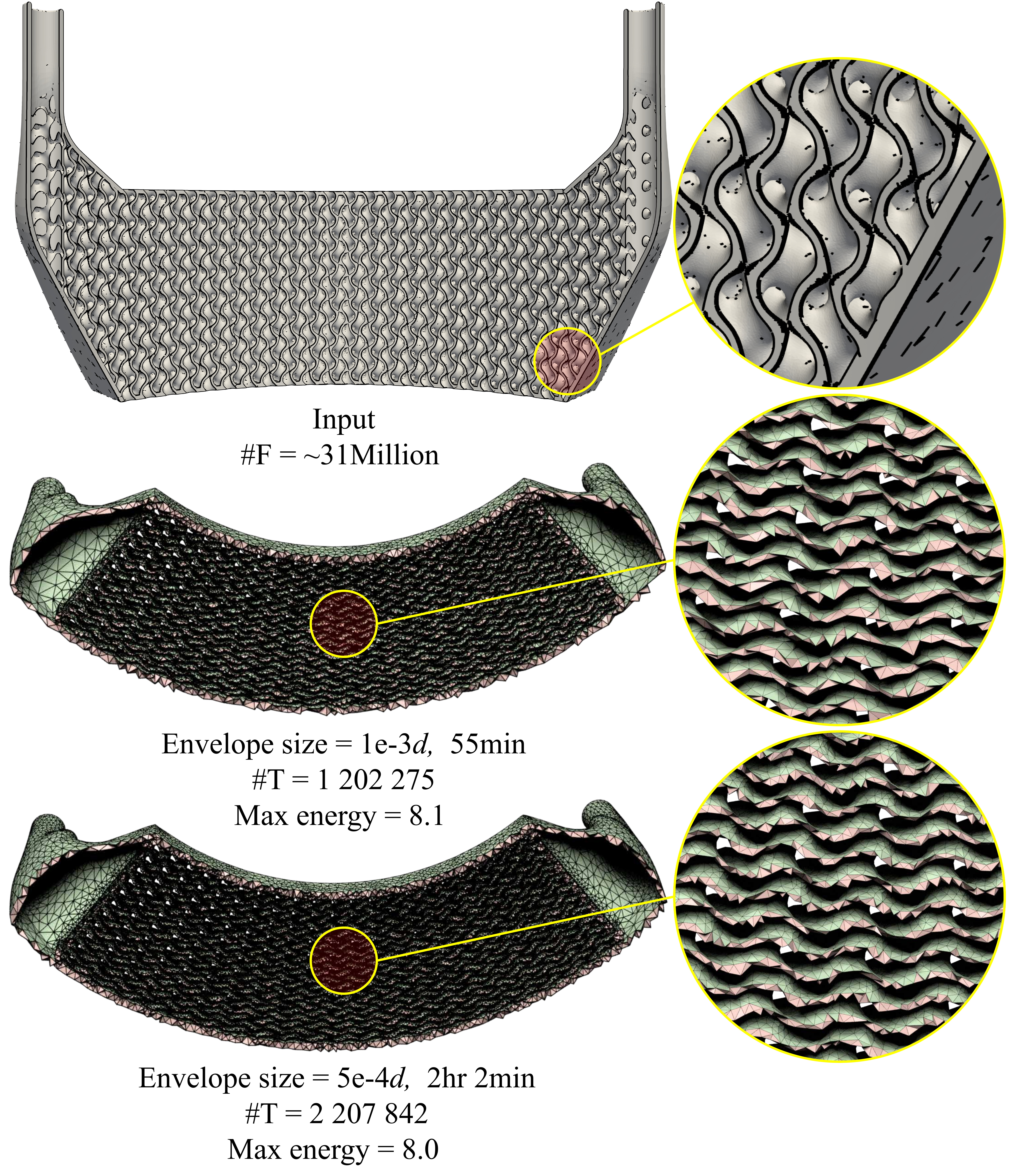}\\
    \caption{Meshing a complex model with 93 million vertices and 31 million faces with different envelope sizes (top). The input mesh contains degenerate triangles and severe self-intersections. Our output tetrahedral meshes are in geometric high quality with either default envelope size (middle) or half envelope size (bottom).}
    \label{fig:ntop}
\end{figure}

\begin{figure}
    \centering\footnotesize
    \includegraphics[width=\linewidth]{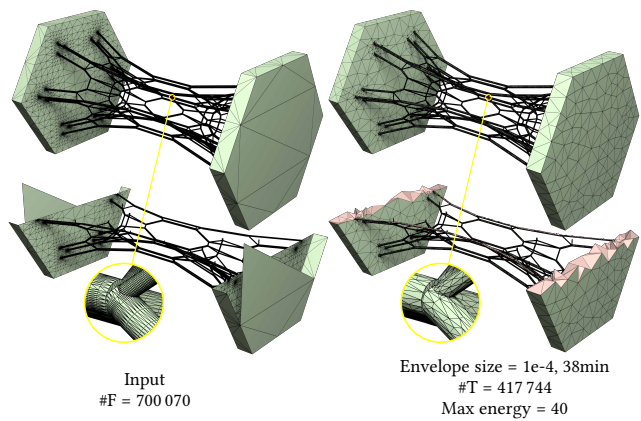}\par
    \parbox{.44\linewidth}{\centering Input\\\#F = 700\,070}\hfill
    \parbox{.44\linewidth}{\centering Envelope size = 1e-4, 38min\\\#T = 417\,744\\Max energy = 40}\par
    \caption{\revision{Example of an architectural application with $80\,999$ self-intersecting faces. The cylinders in the input are intersecting with each other as shown in the closeup. \textsc{fTetWild} successfully cleaned and tetrahedralized this input. Here we stop mesh optimization when maximum energy reaches 50.}}
    \label{fig:architecture}
\end{figure}

We also tested an extremely challenging model coming from an industrial application in additive manufacturing (the part is copyrighted by Velo3D): the design of an exhaust pipe using a volume filled with a structure based on the gyroid triply periodic minimal surface. The model has a multitude of issues introduced during the modeling phase, but it can be cleaned up by our algorithm within 55 minutes (or 122 minutes with  the envelope size decreased by a factor of two), compared to around two weeks of manual labor required by Velo3D's current processing pipeline. Our output mesh (Figure \ref{fig:ntop}) is directly usable for FEA, further editing, or fabrication. As a reference point, the original implementation of TetWild takes 215 minutes with a default envelope size. \revision{Another challenging model we tested contains complex thin structures coming from architecture (Figure~\ref{fig:architecture}). The method in~\cite{masoud20163d,ghomi2018effect} optimizes for the layout of a graph, then replaces the graph edges with cylinders of varying radii. To ensure solidity of the final structure, all cylinders are intersecting as shown in the close up. Although the mesh contains many irregularities, \textsc{fTetWild} successfully meshes the domain into an analysis-ready mesh.}

\begin{figure*}
    \centering\footnotesize
    \includegraphics[width=\linewidth]{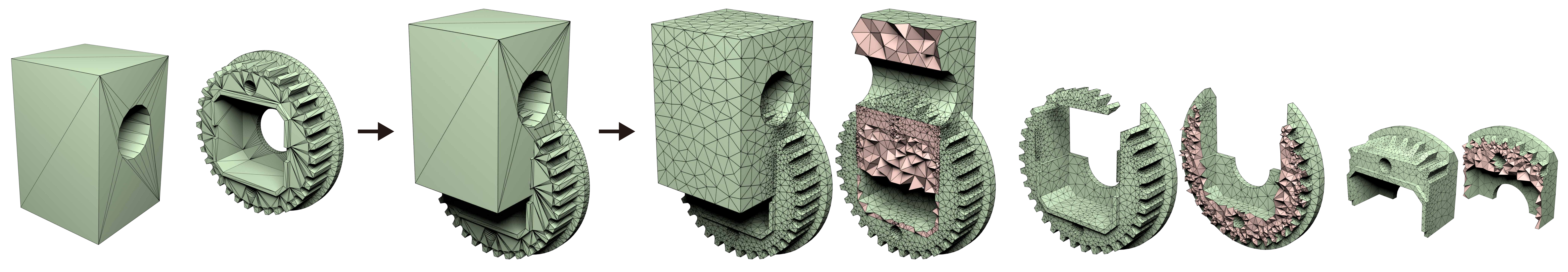}\\
    \parbox{.24\linewidth}{\centering Two objects for Boolean}\hfill
    \parbox{.14\linewidth}{\centering Input\\\#F = 3\,506}\hfill
    \parbox{.6\linewidth}{\centering
        \parbox{.1\linewidth}{~}
        \parbox{.29\linewidth}{\centering Union, 30s\\\#T = 46\,885\\Max energy = 8.0}\hfill
        \parbox{.29\linewidth}{\centering Difference, 31s\\\#T = 25\,768\\Max energy = 8.0}\hfill
        \parbox{.24\linewidth}{\centering Intersection, 33s\\\#T =  10\,347\\Max energy = 7.5}}
    \caption{Three Boolean operations computed on non-manifold, self-intersecting, and non-PWN input surface meshes. The left are two objects for Boolean operation. The middle is the input surface mesh of \textsc{fTetWild}. The right are our output meshes after computing the union, difference, and intersection between the two objects. The average max AMIPS energy of outputs and average time of different operations are with small variance.}
    \label{fig:boolean}
\end{figure*}

\subsection{Mesh Arrangements}\label{sec:bool}

\begin{figure}
    \centering\footnotesize
    \includegraphics[width=\linewidth]{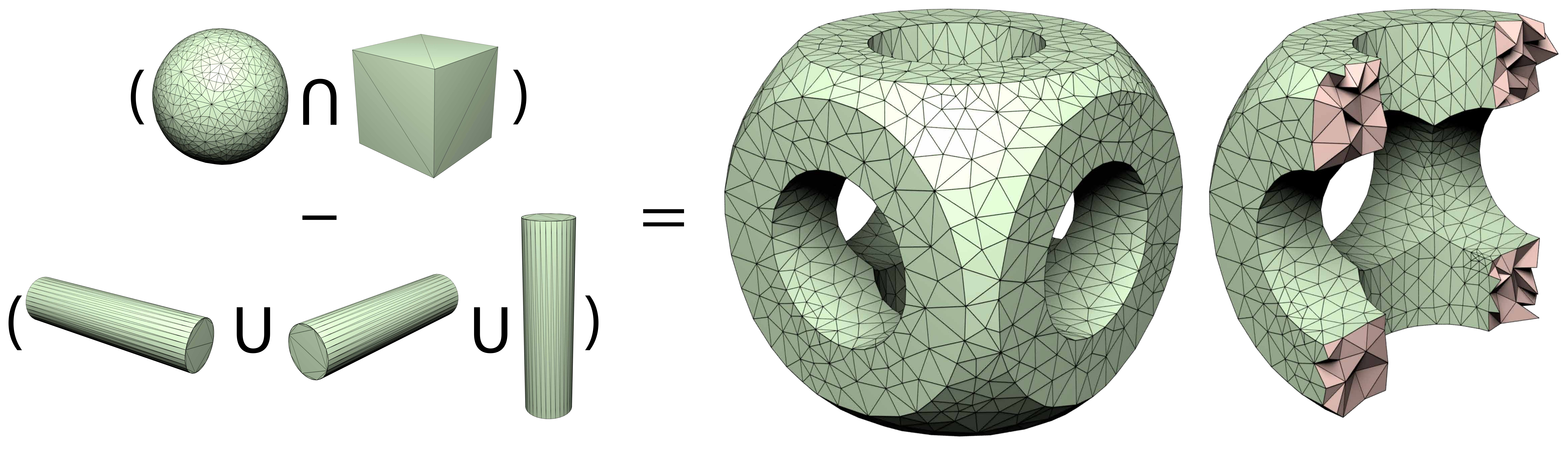}\\
    \caption{\revision{Four Boolean operations among 5 objects. \textsc{fTetWild} takes 34s and products output with \#T = 8\,060 and max energy = 7.2.}}
    \label{fig:boolean1}
\end{figure}

 \citet{Zhou:2016} proposes to compute the arrangement between multiple surfaces using an algorithm to map Boolean operations into simple algebraic expressions involving the winding number of the input surfaces. Their method is robust, but only supports clean PWN surfaces as input.
We propose a simple extension of this algorithm (as explained in Section \ref{sec:met:filtering}) to arbitrary triangle soups. The advantages of our method is evident when the input surfaces come from CAD models containing small gaps or self-intersections: both Mesh Arrangements \cite{Zhou:2016} and CGAL \cite{cgal:hk-bonp3-19a} are unable to perform the operation (since it is not well-defined for non-PWN surface), while \textsc{fTetWild} can compute an approximate (since it allows for an $\epsilon$-deviation from the input surfaces) union, difference, and intersection between them (Figures~\ref{fig:boolean},~\ref{fig:boolean1}), providing robust (but slower) Boolean operations on imperfect geometries. The output is a tetrahedral mesh, which can be useful in downstream applications, and its boundary is a high quality surface triangular mesh.

\begin{figure}
    \centering\footnotesize
    \includegraphics[width=\linewidth]{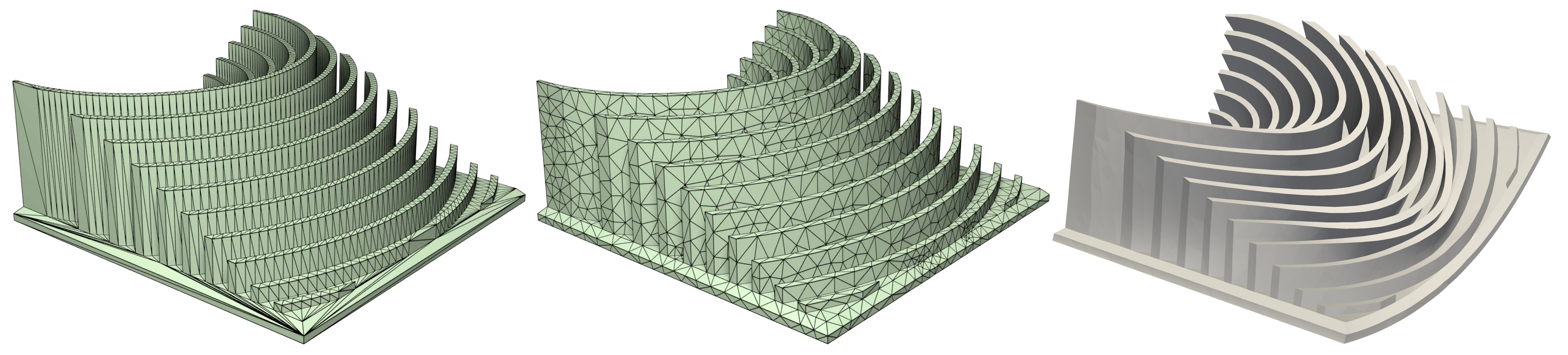}\\[0.4em]
    \parbox{.3\linewidth}{\centering Input\\\#F = 8\,436}\hfill
    \parbox{.3\linewidth}{\centering Time 58s\\\#T = 16\,291\\Max energy = 7.9}\hfill
    \parbox{.3\linewidth}{\centering Elastic deformation}\par
    \caption{\revision{Example of non-linear elastic deformation of a body (right).}}
    \label{fig:fem}
\end{figure}{}

\begin{figure}
    \centering\footnotesize
    \includegraphics[width=\linewidth]{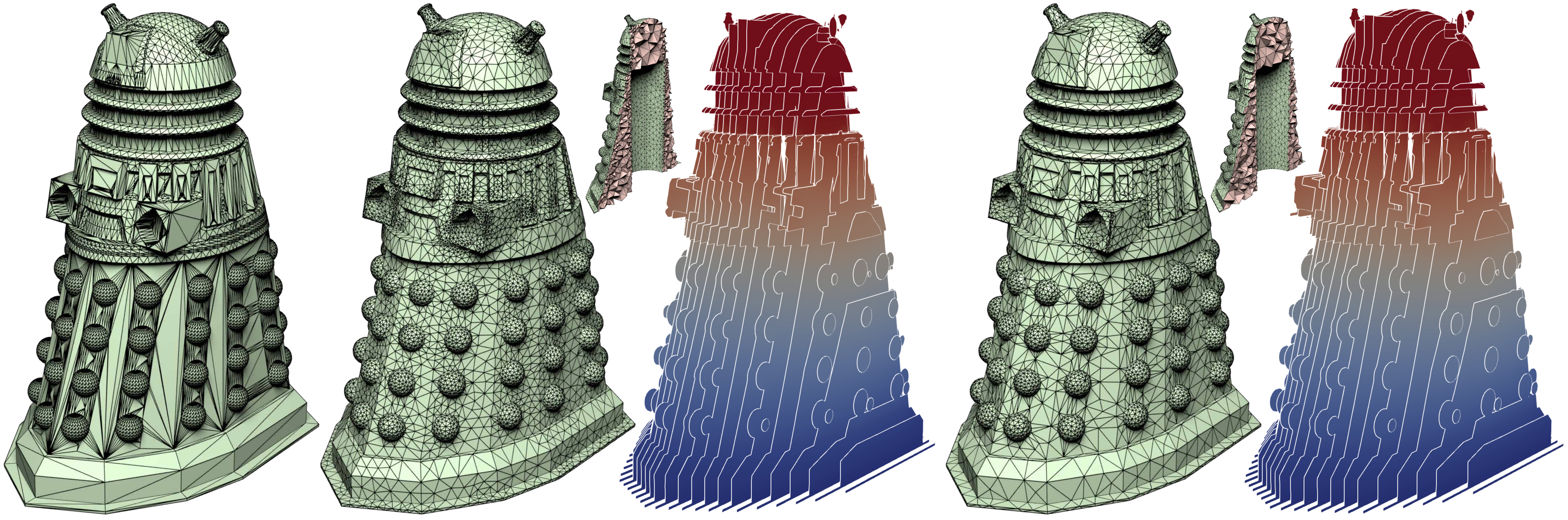}\\[0.4em]
    \parbox{.2\linewidth}{\centering Input\\\#F = 30\,580}\hfill
    \parbox{.38\linewidth}{\centering Max energy $\leq 10$,  107s\\\#T = 90\,438\\Max energy = 8.0}\hfill
    \parbox{.38\linewidth}{\centering $p\leq 4$,  69s\\\#T = 41\,735\\Max energy = 32.4}
    \caption{Two different stopping criteria of our algorithm. The full optimization (middle) improves the mesh to high quality, while using the criterion in~\cite{Schneider:2018:DSA} (right) results in lower mesh quality but faster meshing and smaller mesh size. \revision{The color shows the solution of the volumetric Laplace equation.}}
    \label{fig:p_en_stop}
\end{figure}

\begin{figure}
    \centering\footnotesize
    \includegraphics[width=\linewidth]{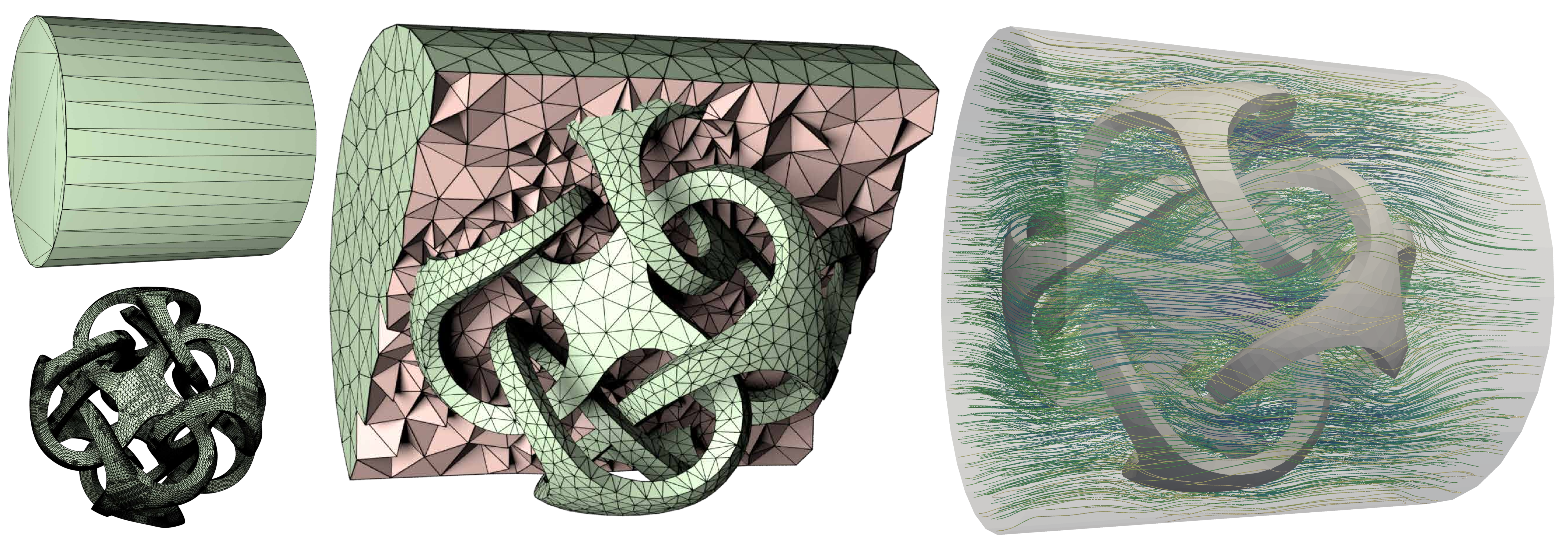}\\[0.4em]
    \parbox{.15\linewidth}{\centering Input\\\#F = 138\,504}\hfill
    \parbox{.4\linewidth}{\centering Time 50s\\\#T = 40\,161\\Max energy = 7.3}\hfill
    \parbox{.4\linewidth}{\centering Streamlines}\par
    \caption{Streamlines of a fluid (right) moving in a cylindrical pipe (left top) with a complicated obstacle (left bottom) in the center. The background mesh (middle) is obtained by subtracting the obstacle from a cylinder using our method.}
    \label{fig:fluid-sim}
\end{figure}

\revision{\subsection{Simulation}}

\revision{The main application of tetrahedral meshing is physical simulations, and the high-quality of our results makes them ideal to be directly used in downstream finite element software (Figure~\ref{fig:fem}).}

Additionally, the recently proposed \emph{a priori} $p$-refinement~\cite{Schneider:2018:DSA} is an ideal fit for our approach when targeting FEM applications, \revision{since \textsc{fTetWild} \emph{always} produces a valid floating-point mesh}. \citet{Schneider:2018:DSA} provides a simple formula to determine the order of each element to compensate for its, possibly bad, shape. We can use this criterion to terminate the mesh optimization early in our algorithm \revision{(thus reducing the meshing time) without affecting the quality of the simulation, Figure~\ref{fig:p_en_stop}.}

\revision{We use the Boolean difference (Section~\ref{sec:bool}) to generate the background mesh} required for simulating the fluid flow on a cylindrical tube containing an obstacle (Figure~\ref{fig:fluid-sim}).

\section{Concluding Remarks}

We introduced \textsc{fTetWild}, a novel robust tetrahedral meshing algorithm for triangle soups which combines the robustness of TetWild with a running time comparable to Delaunay-based methods. The improved performance makes this algorithm suitable not only for applications requiring a volumetric discretization, but also for surface mesh repair and Boolean operations.


\revision{Our current naive parallelization approach shows that our algorithm benefits} from shared-memory parallelization; exploring more advanced parallelization techniques and extending it to distributed computation on HPC clusters are important directions for future work. Our iterative triangle insertion algorithm could be used in dynamic remeshing tasks, potentially allowing to reuse an existing mesh and insert new faces only in regions with high deformation. While conceptually trivial, extending our algorithm to 2D triangle meshing could improve the performance of \cite{Hu:2019}.

Our algorithm uses the conformal AMIPS energy \cite{Rabinovich:2017} to measure and optimize the quality of the tetrahedra. An interesting alternative has been introduced concurrently to our work by \cite{Alexa:2019}: they propose to optimize directly for the Dirichlet energy of the tetrahedralization and show that this measure is effective at removing slivers, while being computationally efficient to evaluate. A comparative study of the two measures would be interesting, and using the Dirichlet energy could lead to further reductions in the running time of our method.
\bibliographystyle{ACM-Reference-Format}
\bibliography{jabbrv.bib,98-bib.bib}

\appendix
\section{A Brief Description of the  TetWild Algorithm}\label{app:tetwild}

\revision{The TetWild algorithm~\cite{Hu:2018} takes a 3D triangle soup as input and generates a tetrahedral mesh that (1) has no inverted or degenerate tetrahedra and (2) contains an approximation of the input surface within a user-defined \emph{$\epsilon$-envelope}. }

\revision{The method starts with an initial background mesh generated using an unconstrained Delaunay tetrahedralization on the input points plus an additional set of evenly-spaced points sampled from a regular grid. These additional points are added to improve the shape of the tetrahedra in the background mesh.}

\revision{
This step generates tetrahedra that might not represent the input faces of the triangle soup: to ensure that they are preserved TetWild uses a Binary Space Partitioning (BSP) subdivision step. Each input triangle is converted into a plane that cuts the tetrahedra of the background mesh. The output of this stage is a polyhedral mesh. To avoid numerical issues and to guarantee that the sub-elements in the polyhedral mesh are convex and non-inverted, TetWild converts the coordinates of the vertices of the initial background mesh into rational numbers and performs all computations using rational numbers. 
}

\revision{Since any convex polyhedron can be trivially subdivided into tetrahedra by adding an additional point in its barycenter, a tetrahedral mesh that exactly preserves input triangles can be naturally obtained after BSP subdivision. However, the vertices of this tetrahedral mesh are represented in rational coordinated. Rounding them to floating point is not simple, since the BSP subdivision introduces badly-shaped tetrahedra which could flip after rounding. TetWild thus increases the quality of the elements using an hybrid optimization procedures that mixes floating point and rational representation.}

\revision{During mesh improvement, the preserved input triangle's faces are tracked and are allowed to move inside the $\epsilon$-envelope. 
The $\epsilon$-envelope limits the tracked surface from deviating from the input further than $\epsilon$.}


\revision{TetWild uses four local operations for mesh improvement: (1) edge splitting, (2) edge collapsing, (3) edge swapping, and (4) vertex smoothing. Every operation is rolled back if the tracked surface leaves the envelope after the operation or if any tetrahedra are inverted, ensuring a valid output. 
Differently from other mesh improvement methods, TetWild uses the 3D conformal AMIPS energy for measuring the geometric quality of the tetrahedra. The AMIPS energy is scaling-invariant and easily differentiable, which boosts these traditional local operations.}

\revision{As the quality of the mesh is improved, the rational coordinates can be gradually rounded into floating points. Theoretically, there might be some unroundable vertices, but it does not occur on the ten thousand models that TetWild has been tested on. 
}

\revision{The final step is the removal of the  tetrahedra outside of the tracked surface. To handle potentially noisy inputs, TetWild computes the winding number of the centroids of all tetrahedra with respect to the tracked surface, and filters out all tetrahedra with centroid's winding-number larger than 0.5. 
}


\section{Example of Unstable AMIPS Energy}\label{app:energy}
\revision{If we compute the 3D AMIPS energy for a tetrahedron with these 4 vertices
\begin{align*}
v_1&=(22.8289586180569, 31.46598870690956, 2.000000016196326)\\
v_2&=(22.83955896584259, 31.46598870610162, 2.000000016081439)\\
v_3&=(22.85206254968259, 31.46598870514861, 2.000000015945925)\\
v_4&=(22.83955896584259, 30.48801551784109, 2.616041190648805)
\end{align*}
we obtain 
\begin{align*}
\mathrm{AMIPS}_{1234}&=5.027711906288343\mathrm{e}10\\
\mathrm{AMIPS}_{2341}&=2.171615254548946\mathrm{e}11\\
\mathrm{AMIPS}_{3412}&=8.865129658843354\mathrm{e}10\\
\mathrm{AMIPS}_{4123}&=7.103076229685612\mathrm{e}10,
\end{align*}
where the subscript indicates the vertex permutations. There are 24 premutations in total and here we pick 4 of them as an example. 
Even if we use the cube of the energy without rational we obtain fluctuations
\begin{align*}
\mathrm{AMIPS}_{1234}^3&=9.401446861483944\mathrm{e}25\\
\mathrm{AMIPS}_{2341}^3&=1.834560196543814\mathrm{e}25\\
\mathrm{AMIPS}_{3412}^3&=1.006679363250288\mathrm{e}26\\
\mathrm{AMIPS}_{4123}^3&=3.462536408842030\mathrm{e}26.
\end{align*}
As reference the correct value computed with rational number is
\[
\mathrm{AMIPS}=1.127562687503913\mathrm{e}11.
\]}


\section{Unused Decompositions of a Tetrahedron}\label{app:secondary}

\begin{figure}
    \centering\footnotesize
    \includegraphics[width=0.65\linewidth]{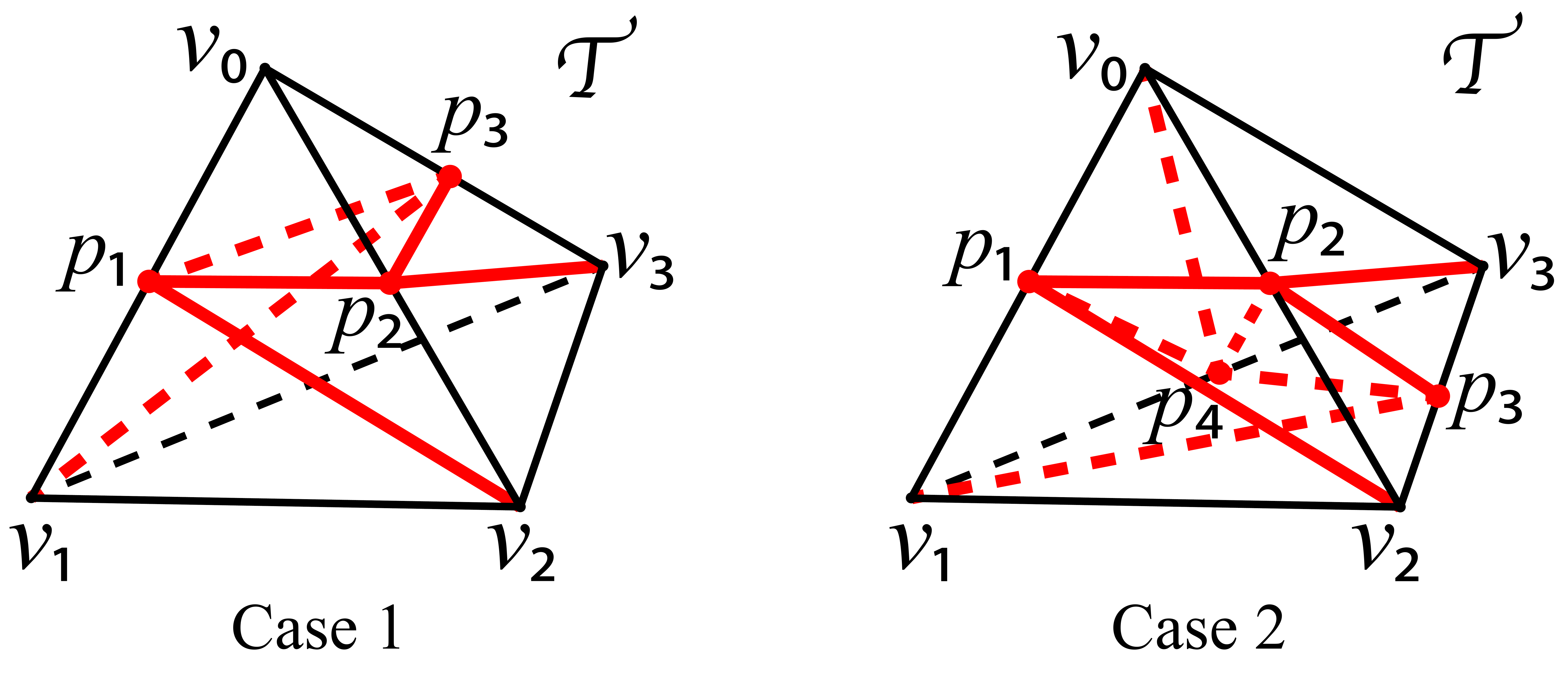}\\
    \caption{\revision{Two unused configurations requiring an additional vertex.}}
    \label{fig:twocases}
\end{figure}

\revision{We enumerated all the possible decompositions of a tetrahedron and discovered two symmetry classes (Figure~\ref{fig:twocases}) of triangulation of faces whose decomposition requires an additional internal vertex. Note that these two cases are never selected by our algorithm (we include them here for completeness), as our rule (Section~\ref{sec:table}) never selects these two cases.}

\revision{
We show that our rule does not select case 1 (Figure~\ref{fig:twocases} left). By contradiction: since the configuration is selected then the edges $[p_1, v_2]$, $[p_2, v_3]$ and $[p_3, v_1]$ are present, thus $v_2>v_1$, $v_3>v_2$, and $v_1>v_3$, according to our rule. Combining these inequalities, the indices of the vertices must satisfy $v_3>v_2>v_1>v_3$, which is impossible. Case 2 (Figure~\ref{fig:twocases} right) is also not selected following a similar argument.}


\section{An Example for Open-Boundary Edge Preservation}\label{app:open}

\begin{figure}
    \centering\footnotesize
    \includegraphics[width=0.85\linewidth]{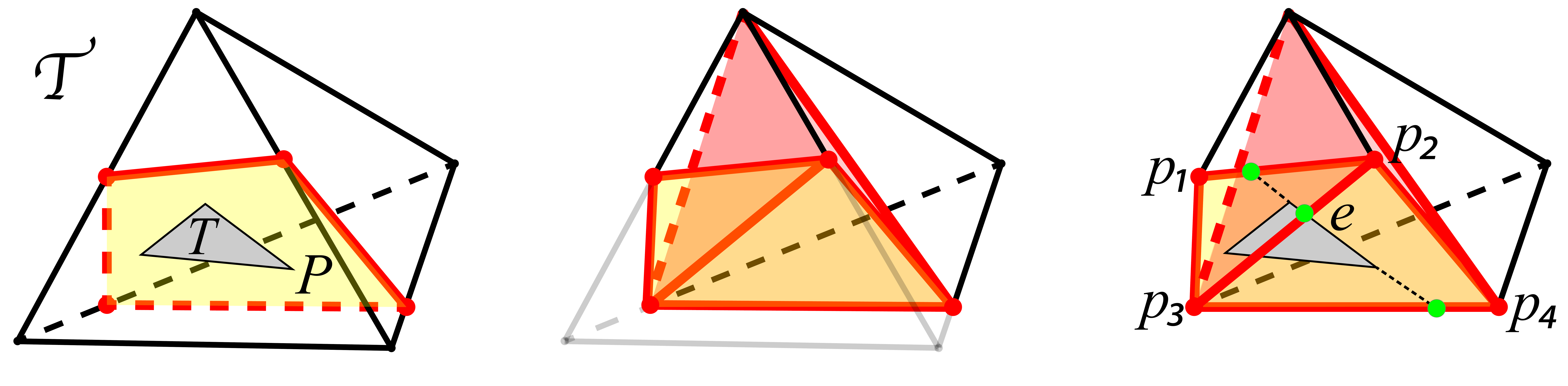}\\
    \parbox{.4\linewidth}{\centering (1)}\hfill
    \parbox{.18\linewidth}{\centering (2)}\hfill
    \parbox{.4\linewidth}{\centering (3)}\par
    \caption{\revision{Example for preserving an open-boundary edge $e$ of triangle $T$. 
    (1) Insert $T$ and $\mathcal{T}_I = \{\mathcal{T}\}$ in this case. (2) The sub-tetrahedra of $\mathcal{T}$ after subdivision. (Only sub-tetrahedra behind $T$ are shown for better visualization.) (3) Inserting edge $e$ and get the intersection points (in green).
    }}
    \label{fig:open-b}
\end{figure}

\revision{
If triangle $T$ is the only inserted triangle and is entirely contained inside a tetrahedron $\mathcal{T}$ (Figure~\ref{fig:open-b}(1)), the intersection of the plane $P$ and $\mathcal{T}$ will be a larger polygon (marked in yellow) containing $T$. In this case, the edges of $T$, which are open-boundary edges, are not preserved. To preserve them, we subdivide the tetrahedra once more.}

\revision{
In Figure~\ref{fig:open-b}(1), $\mathcal{T}$ first get decomposed into sub-tetrahedra (Figure~\ref{fig:open-b}(2)).
Then the faces covering $T$ are $\mathcal{F} = \{[p_1, p_2, p_3]$, $ [p_4, p_2, p_3]\}$ Figure~\ref{fig:open-b}(3). 
The open-boundary edge $e$ and the faces in $\mathcal{F}$ are projected to the best-fitting plane of $p_1, p_2, p_3$, and $p_4$. The intersection points of the projection of $e$ and $\mathcal{T}$ are then computed in 2D and are lifted to 3D (3 green points in Figure~\ref{fig:open-b}(3)). Now there are 3 edges $[p_1, p_2], [p_2, p_3], [p_3, p_4]$ cut into two. We thus subdivide all the neighbouring tetrahedra with the table-based subdivision.}


\section{Changes of Edge-cut Configuration After Snapping}\label{app:aftersnap}

\revision{Table~\ref{table:change} shows all possible edge-cut configurations of a cutting tetrahedron $\mathcal{T}$ after snapping. The final configurations have no more than two vertices which makes the triangulation of $\mathcal{F}$ uniquely defined by the points.
The table includes only the 4 symmetry classes where $\mathcal{T}$ is cut by plane $P$ and contains a face in $\mathcal{F}$ (Figure~\ref{fig:confs} (2)(3)(5)(7)), but excludes the remaining 3 classes where $\mathcal{T}$ is not cut or is just affected by their neighbors (Figure~\ref{fig:confs} (1)(4)(6)).}

\revision{A tetrahedron $\mathcal{T}$ can have at most 3 vertices snapped. If $\mathcal{T}$ has all its 4 vertices within a $\delta$ distance to the $P$, we only snap the 3 vertices closer to $P$.}

\begin{table}
\caption{\revision{Edge-cut configurations of a cutting tetrahedron before and after snapping. Numbers corresponds to the configurations in Figure~\ref{fig:confs}.}}\vspace{-5pt}
\centering\small
 \begin{tabular}{c|c|c|c} 
Before & 1 vertex snapped & 2 vertices snapped & 3 vertices snapped\\ \hline
(2) & (1) & (2) & (1)\\
\hline
(3) & (1)(2) & (1)(2) & (1)\\
\hline
(5) & (1)(3) & (1)(2) & (1)(2) \\
\hline
(7) & (3) & (1)(3) & (1)\\
\end{tabular}\par
\label{table:change}
\end{table}

\end{document}